\documentclass[a4paper,11pt]{article}
\usepackage[margin=2cm]{geometry}
\usepackage{tikz,amsmath, amssymb,bm,color, amsthm}
\usetikzlibrary{positioning, calc,chains,fit,shapes}
\usepackage{enumerate}
\usepackage{comment}
\usepackage{tikz}
\usepackage{graphics}
\usepackage{longtable}
\usepackage{mdframed}
\usepackage{caption}
\usepackage{subcaption}
\usepackage{slashbox}
\usepackage{url}
\usepackage{framed}
\usepackage{array}
\usepackage{tabu}
\usepackage{lscape}
\usepackage{multirow}
\usepackage{ulem}
\usepackage{multicol}
\usepackage{placeins}

\usepackage[noadjust]{cite}
\usepackage[inline]{enumitem}
\usepackage[utf8]{inputenc}
\usepackage[T1]{fontenc}

\usepackage{authblk}

\newlist{mylist}{enumerate*}{1}
\setlist[mylist]{label=(\roman*)}

\mdfsetup{skipabove=2pt,skipbelow=2pt}

\newtheorem{theorem}{Theorem}[section]
\newtheorem{lemma}[theorem]{Lemma}
\newtheorem{observation}[theorem]{Observation}
\newtheorem{corollary}[theorem]{Corollary}
\newtheorem{proposition}[theorem]{Proposition}

\newtheorem{construction}{Construction}

\newcommand{\pname}[1]{\textnormal{\textsc{#1}}}

\newcommand{\FSAT}{\pname{4-SAT$_{\geq 2}$}}

\newcommand{\KSAT}{\pname{$k$-SAT$_{\geq {2}}$}}
\newcommand{\KSATO}{\pname{$k$-SAT}}

\newcommand{\FSATO}{\pname{4-SAT}}

\newcommand{\TSAT}{\pname{3-SAT}}

\newcounter{rowcntr}[table]
\renewcommand{\therowcntr}{\thetable.\arabic{rowcntr}}

\newcolumntype{N}{>{\refstepcounter{rowcntr}\therowcntr}c}

\AtBeginEnvironment{longtabu}{\setcounter{rowcntr}{0}}

\newcounter{rowcntra}[table]
\renewcommand{\therowcntra}{\arabic{rowcntra}}

\newcolumntype{M}{>{\refstepcounter{rowcntra}\therowcntra}c}

\AtBeginEnvironment{tabular}{\setcounter{rowcntra}{0}}

\newcommand{\SCTF}[1]{\pname{SC-to-\ensuremath{\mathcal{F}(#1)}}}
\newcommand{\NPC}{NP-Complete}
\newcommand{\TRUE}{TRUE}

\newcommand{\YES}{YES}
\newcommand{\NO}{NO}
\newcommand{\TETHS}{Further, the problem cannot be solved in time \ensuremath{2^{o(|V(G)|)}}, unless the ETH fails}

\usetikzlibrary{fit}
\usetikzlibrary{
  graphs,
  graphs.standard
}

\usetikzlibrary{decorations.shapes}

\author[1]{Dhanyamol Antony}
\author[2]{Sagartanu Pal}
\author[2]{R. B. Sandeep}

\author[1]{R. Subashini}

\affil[1]{National Institute of Technology Calicut, India

\texttt{\{dhanyamol\_p170019cs,suba\}@nitc.ac.in}}
\affil[2]{Indian Institute of Technology Dharwad, India

\texttt{\{183061001,sandeeprb\}@iitdh.ac.in}}
\title{Cutting a tree with Subgraph Complementation is hard, except for some small trees.}

\date{}
\begin{document}
\maketitle
\begin{abstract}
For a graph property $\Pi$, Subgraph Complementation to $\Pi$ is the problem to find whether there is a subset $S$ of 
vertices of the input graph $G$ such that modifying $G$ by complementing the subgraph induced by $S$ results in a graph satisfying
the property $\Pi$. We prove that the problem of Subgraph Complementation to $T$-free graphs is NP-Complete, for $T$ being a tree,  
except for 41 trees of at most 13 vertices (a graph is $T$-free if it does not contain any induced copies of $T$). 
This result, along with the 4 known polynomial-time solvable cases (when $T$ is a path on at most 4 vertices), leaves behind 37 open cases.
Further, we prove that these hard problems do not admit any subexponential-time algorithms, assuming the Exponential Time Hypothesis.
As an additional result, we obtain that Subgraph Complementation to paw-free graphs can be solved in polynomial-time. 
\end{abstract}

\section{Introduction}
\label{sec:intro}

A graph property is hereditary if it is closed under vertex deletions. 
It is well known that every hereditary property is characterized by a minimal set of forbidden induced subgraphs.
For example, for chordal graphs, the forbidden set is the set of all cycles on at least four vertices, for split graphs, the forbidden set is $\{2K_2, C_4, C_5\}$, 
for cluster graphs it is $\{P_3\}$, and for cographs it is $\{P_4\}$. The study of structural and algorithmic aspects of hereditary graph classes
is central to theoretical computer science. 

A hereditary property is called $H$-free if it is characterized by a singleton set $\{H\}$ of forbidden subgraphs.
Such hereditary properties 
are very interesting for their rich structural and algorithmic properties.
For example, triangle-free graphs could be among the most studied graphs classes.
There is an extensive list of structural studies of $H$-free graphs, for examples, see~\cite{chudnovsky2005structure} for claw-free graphs, 
\cite{DBLP:journals/dam/CorneilLB81} for cographs, and \cite{DBLP:journals/ipl/Olariu88} for paw-free graphs.
There are many important hard problems, such as Independent
Set~\cite{DBLP:journals/jct/Minty80,DBLP:conf/focs/GartlandL20,DBLP:conf/soda/LokshantovVV14,DBLP:conf/sosa/PilipczukPR21,DBLP:journals/talg/GrzesikKPP22}, 
which admit polynomial-time algorithms for $H$-free graphs, for various graphs $H$.

Graph modification problems refer to problems in which the objective is to transform the input graph into a graph with some specific property $\Pi$.
The constraints on the allowed modifications and the property $\Pi$ define a graph modification problem.
For an example, the objective of the Chordal Vertex Deletion problem is to check whether it is possible to 
transform the input graph by deleting at most $k$ vertices so that
the resultant graph is a chordal graph. Graph modification problems, where the target property is $H$-free, have been studied extensively for the last four decades
under various paradigms - exact 
complexity~\cite{AravindSS17,DBLP:journals/siamcomp/Yannakakis81,DBLP:conf/wg/AntonyGPSSS21,DBLP:journals/tcs/AlonS09,DBLP:journals/endm/BrugmannKM09,el1988complexity,DBLP:journals/dam/KomusiewiczU12,DBLP:journals/dam/ShamirST04,sharan2002graph,DBLP:journals/jgt/JelinkovaK14,DBLP:journals/fuin/HageHW03,DBLP:journals/jct/Hayward96,DBLP:journals/dam/Hertz99,kratochvil1992computational}, 
parameterized complexity~\cite{GrammGHN03,drange2015parameterized,DBLP:conf/latin/DrangeDS16}, 
kernelization complexity~\cite{DBLP:journals/jcss/MarxS22,DBLP:journals/disopt/KratschW13,CaiC15incompressibility,cai2012polynomial,DBLP:journals/algorithmica/CaoC12,DBLP:journals/algorithmica/CaoRSY22,GuillemotHPP13,DBLP:journals/tcs/YuanKC21,EibenLS15,DBLP:journals/endm/BrugmannKM09,DBLP:journals/dam/KomusiewiczU12}, 
and approximation complexity~\cite{DBLP:journals/tcs/AlonS09,DBLP:journals/dam/ShamirST04,sharan2002graph,DBLP:journals/toct/BliznetsCKP18}.
We add to this long list by studying the exact complexity of a graph modification problem known as Subgraph Complementation, where the target property is $H$-free.

A \textit{subgraph complement} of a graph $G$ is a graph $G'$ obtained from $G$ by flipping the adjacency of pairs of vertices of a subset $S$ of vertices of $G$.
The operation is known as \textit{subgraph complementation} and is denoted by $G'=G\oplus S$.
The operation was introduced by Kami\'{n}ski et al.~\cite{DBLP:journals/dam/KaminskiLM09} in relation with clique-width of a graph.
For a class $\mathcal{G}$ of graphs,  \textit{subgraph complementation to $\mathcal{G}$} is the problem to check whether there is a set of vertices $S$ in 
the input graph $G$ such that $G\oplus S\in \mathcal{G}$. 
A systematic study of this problem has been started by Fomin et al.~\cite{DBLP:journals/algorithmica/FominGST20}. They obtained
polynomial-time algorithms for this problem for various classes of graphs including triangle-free graphs and $P_4$-free graphs. 
A superset of the authors of this paper studied it further~\cite{DBLP:conf/wg/AntonyGPSSS21} and settled the complexities of this problem (except for a finite number of cases) when $\mathcal{G}$ is $H$-free, 
for $H$ being a complete graph, a path, a star, or a cycle. They proved that subgraph complementation to $H$-free graphs is polynomial-time solvable
if $H$ is a clique, \NPC\ if $H$ is a path on at least 7 vertices, or a star graph on at least 6 vertices, or a cycle on at least 8 vertices. Further, none of these
hard problems admit subexponential-time algorithms, assuming the Exponential-Time Hypothesis. Very recently, an algebraic study of 
\textit{subgraph complementation distance} between two graphs --
the minimum number of subgraph complementations required to obtain one graph from the other -- 
has been initiated by Buchanan, Purcell, and Rombach~\cite{DBLP:journals/combinatorics/BuchananPR22}.

We study subgraph complementation to $H$-free graphs, where $H$ is a tree. We come up with a set $\mathcal{T}$ of 41 trees of at most 13 vertices such that
if $T\notin \mathcal{T}$, then subgraph complementation to $T$-free graphs is \NPC. 
Further, we prove that, these hard problems do not admit subexponential-time algorithms,
assuming the Exponential-Time Hypothesis. 
These 41 trees include some paths, stars, bistars (trees with 2 internal vertices), tristars (trees with 3 internal vertices), and some subdivisions of claw.
Among these, for four paths ($P_\ell$, for $1\leq \ell\leq 4$), the problem is known to be polynomial-time solvable. So, our result leaves behind 
only 37 open cases, which are listed in Figure \ref{table:opencases}.
Additionally, we prove that the problem is hard when $H$ is a 5-connected 
non-self-complementary prime graph with at least 18 vertices. As a separate result, we obtain that the problem can be solved in polynomial-time when $H$ is a paw 
(the unique connected graph on 4 vertices having a single triangle).
\begin{figure}
\begin{center}
\scalebox{0.70}{
\begin{tabular}{| c | c | c | c | c | c | c | c | c | c | c | c | c | c | c |}
\cline{1-3} \cline{5-7} \cline{9-11} \cline{13-15}
\# & \textbf{Name(s)} & \fbox{\textbf{Tree}} & & \# & \textbf{Name(s)} & \textbf{Tree} & & \# & \textbf{Name(s)} & \textbf{\textbf{Tree} } & & \# & \textbf{Name(s)} & \textbf{Tree}\\ \cline{1-3} \cline{5-7} \cline{9-11} \cline{13-15}
1 & \shortstack{$P_5$ \\ $ (T_{1,0,1})$} & \input{figs/open/p5} & &
  11 & $T_{2,2}$ & \input{figs/open/T22} & &
    20 & $T_{1,1,3}$ &\input{figs/open/T113} & &
      29 & $T_{1,3,2}$ & \input{figs/open/T132} \\
      \cline{1-3} \cline{5-7} \cline{9-11} \cline{13-15}
2 & \shortstack{$K_{1,3}$ \\ $ (C_{1,1,1})$} & \input{figs/open/k13} & &
  12 & $T_{2,3}$ & \input{figs/open/T23} & &
    21 & $T_{1,1,4}$ & \input{figs/open/T114} & &
      30 & $T_{1,3,3}$& \input{figs/open/T133} \\
            \cline{1-3} \cline{5-7} \cline{9-11} \cline{13-15}
3 & $K_{1,4}$ & \input{figs/open/K14} & &
  13 & $T_{2,4}$ & \input{figs/open/T24} & &
    22 & $T_{1,1,5}$ & \input{figs/open/T115} & &
      31 & $T_{1,3,4}$  & \input{figs/open/T134} \\
            \cline{1-3} \cline{5-7} \cline{9-11} \cline{13-15}
4 & $C_{1,2,3}$ & \input{figs/open/C123} & &
  14 &$T_{3,3}$ & \input{figs/open/T33} & &
    23 & $T_{1,2,1}$ &\input{figs/open/T121} & &
      32 & $T_{1,3,5}$ & \input{figs/open/T135} \\
      \cline{1-3} \cline{5-7} \cline{9-11} \cline{13-15}
5 & $C_{1,3,3}$ & \input{figs/open/C133} & &
  15 & $T_{3,4}$& \input{figs/open/T34} & &
    24 & $T_{1,2,2}$ &\input{figs/open/T122} & &
      33 & $T_{1,4,1}$ & \input{figs/open/T141} \\
      \cline{1-3} \cline{5-7} \cline{9-11} \cline{13-15}
6 & $C_{2,2,2}$ & \input{figs/open/C222} & &
  16 &$T_{4,4}$ & \input{figs/open/T44} & &
    25 & $T_{1,2,3}$ &\input{figs/open/T123} & &
      34 & $T_{1,4,2}$  & \input{figs/open/T142} \\
      \cline{1-3} \cline{5-7} \cline{9-11} \cline{13-15}
7 & $C_{2,2,3}$ & \input{figs/open/C223} & &
  17 & \shortstack{$T_{1,0,2}$ \\ $ (C_{1,1,3})$} & \input{figs/open/T102} & &
    26 & $T_{1,2,4}$ &\input{figs/open/T124} & &
      35 & $T_{1,4,3}$ & \input{figs/open/T143} \\
      \cline{1-3} \cline{5-7} \cline{9-11} \cline{13-15}
8 & \shortstack{$T_{1,2}$ \\ $ (C_{1,1,2})$} & \input{figs/open/T12} & &
  18 & \shortstack{$T_{1,1,1}$ \\ $ (C_{1,2,2})$} & \input{figs/open/T111} & &
    27 & $T_{1,2,5}$ &\input{figs/open/T125} & &
      36 & $T_{1,4,4}$ & \input{figs/open/T144} \\
      \cline{1-3} \cline{5-7} \cline{9-11} \cline{13-15}
9 & $T_{1,3}$ & \input{figs/open/T13} & &
  19 & $T_{1,1,2}$& \input{figs/open/T112} & &
    28 & $T_{1,3,1}$  &\input{figs/open/T131} & &
      37 & $T_{1,4,5}$ & \input{figs/open/T145} \\
      \cline{1-3} \cline{5-7} \cline{9-11} \cline{13-15}
10 & $T_{1,4}$ & \input{figs/open/T14} \\
      \cline{1-3} 
\end{tabular}}
\end{center}
\caption{The trees $T$ for which the complexity of \SCTF{T} is open}
\label{table:opencases}
\end{figure}

\section{Preliminaries}
\label{sec:prelim}

In this section, we provide various definitions, notations, and terminologies used in this paper.

\textbf{Graphs}.
    For a graph $G$, the vertex set and edge set are denoted by $V(G)$ and $E(G)$ respectively. 
    A graph $G$ is \textit{$H$-free} if it does not contain $H$ as an induced subgraph. 
    By $\mathcal{F}(H)$ we denote the class of $H$-free graphs.
    The vertex connectivity, $\mathcal{K}(G)$, of a graph $G$ is 
    the minimum number of vertices in $G$ whose removal either causes $G$  disconnected or reduces $G$ to a graph with only one vertex.
    A graph $G$ is said to be $k$-connected, if $\mathcal{K}(G)\geq k$. 
    By $K_n$, $nK_1$, $K_{1,n-1}$, $C_n$, and $P_n$, we denote the complete graphs, empty graphs, star graphs, cycles, and paths on $n$ vertices respectively.

    A graph $G$ which is  isomorphic to its complement $\overline{G}$ is called a \textit{self-complementary} graph. 
    If $G$ is not isomorphic to $\overline{G}$, then it is called \textit{non-self-complementary}. 
    The \textit{join} of two graphs $G$ and $H$, denoted by $G\times H$, is a graph in which each vertex in $G$ is adjacent to all vertices in $H$.
    By $G+H$, we denote the  \textit{disjoint union} of two graphs $G$ and $H$. 
    Similarly by  $rG$, we denote the disjoint union of $r$ copies of a graph $G$.
    By $G[H]$, we denote the graph obtained from $G$ by replacing each vertex of $G$ with $H$. That is, 
    $V(G[H]) = V(G)\times V(H)$, and $E(G[H]) = \{(u,v)(u',v')| (u,u')\in E(G)\ \text{or}\ (u=u' \text{and}\ (v,v')\in E(H))\}$.  
    For a subset $X$ of vertices of $G$, by $G-X$ we denote the graph obtained from $G$ by removing the vertices in $X$.

    The open neighborhood of a vertex $v\in V(G)$, denoted by $N(v)$, is the set of all the vertices adjacent to $v$, i.e., 
    $N(v):=\{w $ $\mid$ $vw \in E(G)\}$, and the closed neighborhood of $v$, denoted by $N[v]$, 
    is defined as $N(v) \cup \{v\}$. 
    A pair of non-adjacent vertices in a graph $G$ are called \textit{false-twins}, if they have the same neighborhood in $G$. 
    Let $u$ be a vertex and $X$ be a vertex subset of $G$. By $N_X(u)$ and $N_{\overline{X}}(u)$, we  
    denote the neighborhood of $u$ inside the sets $X$ and $V(G) \setminus X$, respectively. 
    We extend the notion of adjacency to sets of vertices as: two sets $A$ and $B$ of vertices of 
    $G$ are \textit{adjacent} (resp., \textit{non-adjacent}) if each vertex of $A$ is adjacent (resp., non-adjacent) 
     to each vertex of $B$. 
     We say that a graph $H$ is obtained from $H'$ by \textit{vertex duplication}, if $H$ is obtained from $H'$ by replacing each vertex $v_i$ in $H'$
     by an independent set of size $r_i\geq 1$.

    A \textit{tree} is a connected acyclic graph, and a disjoint union of trees is called a \textit{forest}. 
    The \textit{internal tree} $T'$ of a tree $T$ is a tree obtained by removing all the leaves of $T$. 
    The \textit{center} of a star graph $K_{1,x}$ is the vertex which is connected to all the leaves of $K_{1,x}$.
    A \textit{bistar} graph $T_{x,y}$, for $x\geq 1$ and $y\geq 1$, is a graph obtained by making $a$ and $b$ adjacent, 
    where $a$ and $b$ are the centers of two star graphs $K_{1,x}$ and $K_{1,y}$ respectively. Here $a$ is the $x$-center 
    (the vertex adjacent to $x$ leaves)  and $b$ is the $y$-center (the vertex adjacent to $y$ leaves) 
    of  $T_{x,y}$. 
    The bistar graph $T_{1,1}$ is isomorphic to $P_4$.
    Similarly,  \textit{tristar} graph $T_{x,y,z}$, for $x\geq 1$, $y\geq 0$ and $z\geq 1$, 
    is a graph obtained by joining the centers $a,b,$ and $c$ of three star graphs $K_{1,x}$, $K_{1,y}$, 
    and $K_{1,z}$ respectively in such a way that $\{a,b,c\}$ induces a $P_3$ with $b$ as the center. 
    The tristar graph $T_{1,0,1}$ is isomorphic to $P_5$.
    A \textit{subdivision of claw}, denoted by $C_{x,y,z}$ for $1\leq x\leq y\leq z$, 
    is a graph obtained from the claw, $K_{1,3}$, 
    by subdividing its three edges $x-1$ times, $y-1$ times, and $z-1$ times respectively. 
    The subdivision of claw $C_{1,1,1}$ is isomorphic to the claw.
    Similarly, a subdivision of a star $K_{1,a}$, for $a\geq 3$, is denoted by $C_{x_1,x_2,\ldots,x_{a}}$, where $1\leq x_1\leq x_2\leq\ldots\leq x_a$.

    \textbf{Modular decomposition.}
    A vertex subset $X$ of $G$ is a \textit{module}
    if $N_{\overline{X}}(u) = N_{\overline{X}}(v)$ for all 
    $u,v \in X$. The \textit{trivial} modules of a graph $G$ are   $\emptyset , V(G),$ and all the singletons $\{ v\}$ for $v \in V(G)$. 
    A graph is \textit{prime} if it has at least 3 vertices and all its modules are trivial, and \textit{nonprime} otherwise.
    A nontrivial module $M$ is a \textit{strong module} of a graph $G$ if for every other module $M'$ in $G$, if $M\cap M'\neq \emptyset$, then 
    either $M\subseteq M'$ or $M'\subseteq M$. 
    A module which induces an independent set is called \textit{independent module} and a module which induces a clique is called \textit{clique module}.
    Let $G$ be a nonprime graph such that both $G$ and $\overline{G}$ are connected graphs. 
    Then there is a unique partitioning $\mathcal{P}$ of $V(G)$ into maximal strong modules. 
    The quotient graph $Q_G$ of $G$ has one vertex for each set in $\mathcal{P}$ and two vertices in $Q_G$ are adjacent if and only if the corresponding 
    modules are adjacent in $G$. The modular decomposition theorem due to Gallai~\cite{gallai1967transitiv} says that $Q_G$ is a prime graph.
    We refer to \cite{DBLP:journals/csr/HabibP10} for more details on modular decomposition and related concepts.

\textbf{Boolean satisfiability problems}.
In a \TSAT\ formula, every clause contains exactly three literals of distinct variables and the  objective of the \TSAT\ problem is to 
find whether there exists a truth assignment which assigns \TRUE\ to at least one literal per clause. The problem is among the first known 
\NPC\ problems.
The Exponential-Time Hypothesis (ETH) and the Sparsification Lemma imply that \TSAT\ cannot be solved in subexponential-time, i.e., 
in time $2^{o(n+m)}$, where $n$ is 
the number of variables and $m$ is the number of clauses in the input formula.
To prove that a problem does not admit a subexponential-time algorithm, it is sufficient to obtain a linear reduction from a problem known not to admit
a subexponential-time algorithm, where a linear reduction is a polynomial-time reduction
in which the size of the resultant instance is linear in the size of the input instance.
All our reductions are trivially linear and we may not explicitly mention the same.
We refer to the book~\cite{DBLP:books/sp/CyganFKLMPPS15} for a detailed description of these concepts.

In a \KSATO\ formula, every clause contains exactly $k$ literals. The objective of the \KSAT\ problem is to  
find whether there is a truth assignment for the input \KSATO\ formula such that at least two literals per clause are assigned \TRUE.
For every $k\geq 4$, there are two simple linear reductions from \TSAT\ to \FSAT\ and then to \KSAT\ to prove the hardness of \KSAT.
Replace every clause $(\ell_1\lor \ell_2\lor \ell_3)$ in the input $\Phi$ of \TSAT\ by a clause 
$(\ell_1\lor \ell_2\lor \ell_3\lor x_1)$ - this makes sure that
$\Phi$ is satisfiable if and only if there is a truth assignment which assigns \TRUE\ to at least two literals per clause of the new formula.
A linear reduction from \FSAT\ to \KSAT\ is also trivial: Replace every clause $(\ell_1\lor \ell_2\lor\ell_3\lor\ell_4)$ in the input $\Phi$ of \FSAT\ 
by $2^{k-4}$ clauses each of them contains $\ell_1, \ell_2, \ell_3,\ell_4$, and either the positive literal or the negative literal of $k-4$
new variables $x_1, x_2, \ldots, x_{k-4}$. The $2^{k-4}$ clauses are to make sure that all combinations of the negative and positive literals of the
new variables are present which makes sure that $\Phi$ is satisfiable (with two true literal per clause) 
if and only if the new formula is satisfiable (with two true
literals per clause). Since $k$ is a constant, this reduction is a linear reduction.

\begin{proposition}[folklore]
    \label{pro:ksat}
    For $k\geq 4$,
    \KSAT\ is \NPC. 
    Further, the problem cannot be solved in time $2^{o(n+m)}$, assuming the ETH.
\end{proposition}

    By $G\oplus S$, for a graph $G$ and $S\subseteq V(G)$, we denote the graph obtained from $G$ by flipping the adjacency of pairs of vertices in $S$.
    The problem that we deal with in this paper is given below.
    
\begin{mdframed}
  \textbf{\SCTF{H}\ }: Given a graph $G$, find whether there is a set $S\subseteq V(G)$ such that $G\oplus S$ is $H$-free.
\end{mdframed}

We make use of the following known results.

\begin{proposition}[\cite{DBLP:conf/wg/AntonyGPSSS21}]
    \label{pro:path}
    Let $T$ be a path on at least 7 vertices. Then \SCTF{T} is \NPC. \TETHS.
\end{proposition}

\begin{proposition}[\cite{DBLP:conf/wg/AntonyGPSSS21}]
    \label{pro:star}
    Let $T$ be a star on at least 6 vertices. Then \SCTF{T} is \NPC. \TETHS.
\end{proposition}

We say that two problems $A$ and $B$ are linearly equivalent, if there is a linear reduction from $A$ to $B$ 
and there is a linear reduction from $B$ to $A$.

\begin{proposition}[\cite{DBLP:conf/wg/AntonyGPSSS21}]
    \label{pro:complement}
    \SCTF{H} and \SCTF{\overline{H}} are linearly equivalent.
\end{proposition}

\section{Reductions for general graphs}
\label{sec:general}

In this section, we introduce two reductions which will be used in the next section to prove hardness for \SCTF{H}, when $H$ is a tree. 
We believe that these reductions will be useful in an eventual dichotomy for the problem for general graphs $H$.
The first reduction is a linear reduction from \SCTF{H'} to \SCTF{H} where $H$ is obtained from $H'$ by vertex duplication.
The second reduction proves that for every 5-connected non-self-complementary prime graph $H$ with a clique or independent set of size $4$, 
\SCTF{H} is \NPC\ and does not admit a subexponential-time algorithm, assuming the ETH.
\subsection{Graphs with duplicated vertices}
\label{sub:graphs with duplicated vertices}
Here, with the help of a linear reduction, we prove that the hardness results for a prime graph $H'$ \textcolor{blue}{}{translate} to that for $H$,
where $H$ is obtained from $H'$ by vertex duplication.
\begin{lemma}
   \label{lem:duplication}
   Let $H'$ be a prime graph with vertices $V(H') = \{v_1,v_2,\ldots, v_t\}$. 
Let $H$ be a graph obtained from $H'$ by replacing each vertex $v_i$ in $H'$ by 
an independent set $I_i$ of size $r_i$, for some integer $r_i\geq 1$. 
Then there is a linear reduction from \SCTF{H'} to \SCTF{H}.
\end{lemma}

\begin{figure}[!htbp]
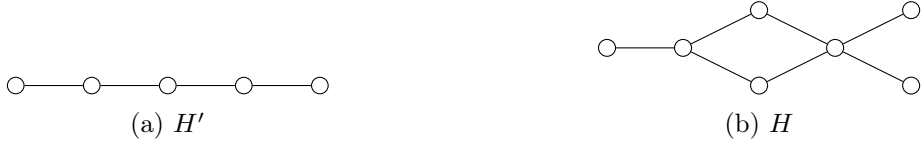

 \centering
    \begin{subfigure}[b]{0.45\textwidth}
          \centering
          {\input{figs/hardness/egG1-duplicated}}  
          \caption{$H'$}
          \label{fig:H'A}
     \end{subfigure}
     \begin{subfigure}[b]{0.450\textwidth}
          \centering
          {\input{figs/hardness/egG-duplicated}}  
          \caption{$H$}
          \label{fig:HrB}
     \end{subfigure}
     \caption{An example of $H'$ and $H$. Here, $r_1=r_2=r_4=1, r_3=r_5=2$, and $r=2$, assuming an ordering of vertices of $H'$ from left to right.}
    \label{fig:vertex duplicated eg}
\end{figure}

Let $H'$ and $H$ be graphs mentioned in Lemma~\ref{lem:duplication}.
Let $r$ be the maximum integer among the $r_i$s, i.e., $r=\max_{i=1}^{i=t}{r_i}$.
We note that $H'$ is the quotient graph of $H$. See Figure~\ref{fig:vertex duplicated eg} for an example.

\begin{figure}[!htbp]
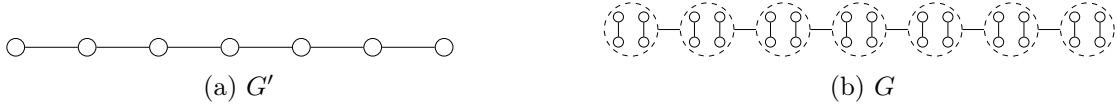

    \begin{subfigure}[b]{0.450\textwidth}
          \centering
         \resizebox{0.8\textwidth}{!}{\input{figs/hardness/G1-leaves_duplicated} } 
          \caption{$G'$}
          \label{fig:A}
     \end{subfigure}
     \begin{subfigure}[b]{0.50\textwidth}
          \centering
           \resizebox{0.8\textwidth}{!}{\input{figs/hardness/G-leaves_duplicatd}}  
          \caption{$G$}
          \label{fig:B}
     \end{subfigure}

     \caption{{An example of Construction \ref{cons vertex duplicated} for a graph $G'$ isomorphic to $P_7$, and for an interger $r=2$. 
     The lines connecting two circles (bold or dashed) 
     indicate that the vertices corresponding to that circles are adjacent. 
     }}
      \label{fig vertex duplicated}
\end{figure}

\begin{construction}
\label{cons vertex duplicated}
     Given a graph $G'$ and an integer $r\geq 1$, the graph $G$ is constructed from $G'$ as follows: 
     for each vertex $u$ of $G'$, replace $u$ with a set $W_u$ which induces an $rK_r$.
    The so obtained graph is $G$ (see Figure ~\ref{fig vertex duplicated} 
    for an example). 
\end{construction}

\begin{lemma}\label{lem vertex duplicated if}
   If $G' \oplus S' \in \mathcal{F}(H')$ for some $S' \subseteq V(G')$, then   $G \oplus S \in \mathcal{F}(H)$, 
   where $S$ is the union of vertices in $W_u$ for every vertex $u\in S'$.
\end{lemma}

\begin{proof}
    Let an $H$ be induced by $A$ (say) in $G\oplus S$. 
    Recall that $G$ is constructed by replacing each vertex $u$ in $G'$ with a module $W_u$ which induces an $rK_r$.
    If $A\subseteq W_u$ for some vertex $u$ in $G'$, then $H$ is an induced subgraph of either $rK_r$ (if $u\notin S'$) or $\overline{rK_r}$ (if $u\in S'$).
    Then $H'$, the quotient graph of $H$, is either an independent set or a complete graph. This is not true as $H'$ is a prime graph.
    Therefore, $A$ has nonempty intersection with more than one $W_u$s. For a vertex $u$ in $G'$, either $W_u$ 
    is a subset of $S$ (if $u\in S'$) or $W_u$ has empty intersection with $S$ (if $u\notin S'$).
    Therefore, if $A$ has nonempty intersection with $W_u$, then $A\cap W_u$ is a module of the $H$ induced by $A$. 
    Therefore, $A\cap W_u\subseteq I_i$ for some $1\leq i\leq t$.
    Let $U_i$ be the set of vertices $u$ in $G'$ such that $I_i$ (in the $H$ induced by $A$) has a nonempty intersection with $W_u$.
    Arbitrarily choose one vertex from $U_i$. Let $A'$ be the set of such chosen vertices for all $1\leq i\leq t$.
    We claim that $A'$ induces an $H'$ in $G'\oplus S'$.
    Let $u_i$ and $u_j$ be the vertices chosen for $I_i$ and $I_j$ respectively, for $i\neq j$.
    Since $A\cap W_{u_i}\subseteq I_i$ and $A\cap W_{u_j}\subseteq I_j$, and $i\neq j$, we obtain that $u_i\neq u_j$.
    It is enough to prove that $u_i$  and $u_j$ are adjacent in $G'\oplus S'$
    if and only if $v_i$ and $v_j$ are adjacent in $H'$. If $u_i$ and $u_j$ are adjacent in $G'\oplus S'$, then $W_{u_i}$ and $W_{u_j}$
    are adjacent in $G\oplus S$. 
    This implies that $I_i$ and $I_j$ are adjacent in $H$. Hence $v_i$ and $v_j$ are adjacent in $H'$.
    For the converse, assume that $v_i$ and $v_j$ are adjacent in $H'$. This implies that $I_i$ and $I_j$ are adjacent in $H$.
    Therefore, $W_{u_i}$ and $W_{u_j}$ are adjacent in $G\oplus S$. Hence $u_i$ and $u_j$ are adjacent in $G'\oplus S'$. This completes the proof.
\end{proof}

\begin{lemma}\label{lem vertex duplicated onlyif}
   If $G \oplus S \in \mathcal{F}(H)$ for some $S \subseteq V(G)$, then   $G' \oplus S' \in \mathcal{F}(H')$, 
   where $S'$ is a  subset of vertices of $G'$ obtained in such a way that whenever all vertices of a $K_r$ 
   from a module $W_u$ (which induces an $rK_r$) are in $S$, then the corresponding vertex $u$ in $G'$ is included in $S'$.
\end{lemma}

\begin{proof}
    Suppose $G' \oplus S'$ contains an $H'$ induced by a set $A'=\{v_1, v_2, \ldots, v_t\}$.
    If a vertex $u$ in $G'$ is in $S'$, then all vertices of a $K_r$ from $W_u$ in $G$  are in $S$. 
    Therefore, there is an independent set of size $r$ in $W_u\cap S$ in $G\oplus S$.
    Similarly, if $u\notin S'$, then there is an independent set of size $r$ in $W_u\setminus S$ in $G\oplus S$ formed by one vertex, which is not in $S$, 
    from each copy of $K_r$ in $W_u$ 
    which is not in $S$. We construct $A$ as follows: for each vertex $v_i\in A'$, if $v_i\in S'$, include in $A$ an independent set $I_i\subseteq W_{v_i}\cap S$
    such that $|I_i|=r_i$, 
    and if $v_i\notin S'$, include in $A$ an independent set $I_i\subseteq W_{v_i}\setminus S$ such that $|I_i| = r_i$.
    We claim that $A$ induces an $H$ in $G\oplus S$. Note that each chosen $I_i$ is a module in $G\oplus S$.
    Since $I_i\subseteq S$ if and only if $v_i\in S'$, we obtain that $I_i$ and $I_j$ are adjacent in $G\oplus S$ 
    if and only if $v_i$ and $v_j$
    are adjacent in the $H'$ induced by $A'$.
    This completes the proof.
\end{proof}

Lemma~\ref{lem:duplication} follows directly from Lemma~\ref{lem vertex duplicated if} and \ref{lem vertex duplicated onlyif}.
When the lemma is applied on trees, we get the following corollary. We note that the quotient tree $Q_T$ of a tree is prime
if and only if $T$ is not a star graph - by our definition, a prime graph has at least 3 vertices.

\begin{corollary}
\label{cor:tree-duplication}
Let $T$ be a tree which is not a star graph, and let $Q_T$ be its quotient tree. Then there is a linear reduction from \SCTF{Q_T}\ to \SCTF{T}.
\end{corollary}

\subsection{5-connected graphs}\label{sub:5-conn}
Here, we obtain hardness results for \SCTF{H}, where $H$ is a 5-connected graph satisfying some additional constraints.
\begin{theorem}
\label{thm:5-conn}
    Let $H$ be a 5-connected non-self-complementary prime graph with an independent set of size 4 
    or with a clique of size 4. 
    Then \SCTF{H} is \NPC. \TETHS.
\end{theorem}

We have the following corollary from the fact that the Ramsey number $R(4,4) = 18$.
\begin{corollary}
\label{cor:5-connected}
Let $H$ be a 5-connected non self-complementary prime graph with at least 18 vertices. Then \SCTF{H}\ is \NPC. \TETHS.
\end{corollary}

Let $H$ be a 5-connected graph satisfying the constraints mentioned in Theorem~\ref{thm:5-conn}.
Let $H$ have $t$ vertices and let $V'\subseteq V(H)$ induce either a $K_4$ or a $4K_1$ in $H$.
We use Construction~\ref{cons 5-connected} for a reduction from \FSAT\ to prove Theorem~\ref{thm:5-conn}.

\begin{construction}
\label{cons 5-connected}
     Let $\Phi$ be a \FSATO\ formula with $n$ variables $X_1, X_2, \cdots, X_n$, and $m$ clauses $C_1, C_2, \cdots,$ $C_m$. 
    We construct the graph $G_{\Phi}$ as follows. 
    \begin{itemize}
    \item For each variable $X_i$ in $\Phi$, the variable gadget also named as $X_i$ consists of the union of
    two special sets $X_{i1}=\{x_i\}$ and $X_{i2} = \{\overline{x_i}\}$, and $t-2$ other sets $X_{i3}, X_{i4} \dots X_{it}$
    such that each $X_{ij}$, for $3\leq j\leq t$ induces an $\overline{H}$. 
    Make the adjacency between these $X_{ij}$s in such a way that taking one vertex each from these sets induces an $H$, where $X_{i1}$
    and $X_{i2}$ correspond to two non-adjacent vertices, if $V'$ forms a $K_4$, and correspond to two adjacent vertices, if $V'$ forms 
    a $4K_1$. 
    If $V'$ forms a clique then add an edge between $X_{i1}$ and $X_{i2}$, and if $V'$ forms an independent set, then remove the edge between $X_{i1}$ and $X_{i2}$.
    The vertices $x_i$s and $\overline{x}_i$s are called literal vertices denoted by a set $L$, which induces a clique, if $V'$ is a clique, 
    and induces an independent set, if $V'$ is an independent set.
    \item  For each clause $C_i$  of the form  $(\ell_{i1}\lor \ell_{i2}\lor \ell_{i3}\lor \ell_{i4})$ in $\Phi$, the clause gadget also named as $C_i$ consists of $t-4$ copies of $\overline{H}$ denoted by $C_{ij}$, for $1\leq j\leq (t-4)$. 
    Let the four vertices introduced (in the previous step) for the literals $\ell_{i1}, \ell_{i2}, \ell_{i3}$, and $\ell_{i4}$ be denoted by  
    $L_i=\{y_{i1},y_{i2},y_{i3},y_{i4}\}$. 
    The adjacency among each of these  $C_{ij}$s and the literal vertices $L_i$ is in such a way that, 
    taking one vertex from each $C_{ij}$s and the vertices in $L_i$ induces an $H$. 
   
    \end{itemize}
    This completes the construction.\end{construction}

\begin{figure}[!htbp]
    \centering
    \input{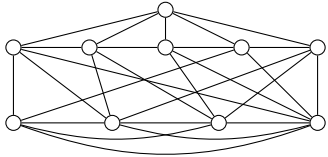}  
    \caption{An example of a 5-connected non-self-complementary prime graph with a $K_4$ (formed by the lower four vertices)}
   \label{fig: hfree 5-con k4 eg}
  \end{figure}

\begin{figure}[ht]
\centering
\input{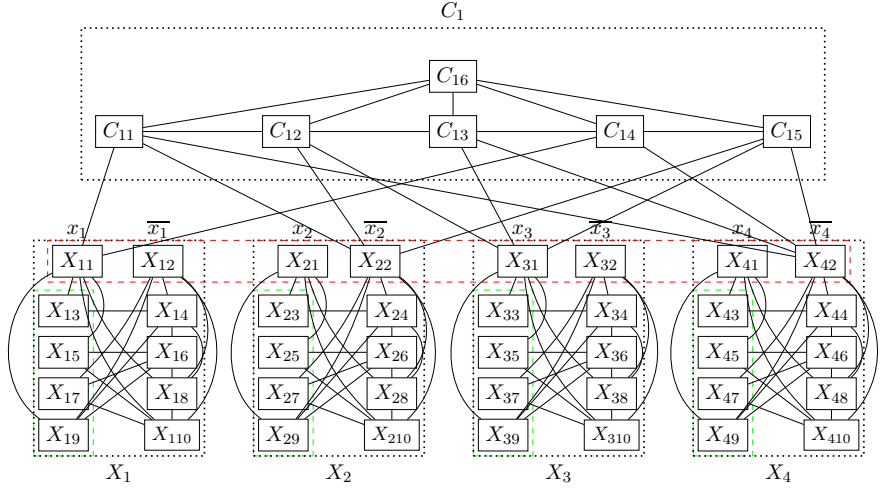}
  \caption{An example of Construction \ref{cons 5-connected} for the formula $\Phi = C_1$ 
  where $C_1=x_1\lor \overline{x_2}\lor x_3\lor \overline{x_4}$ corresponding to the graph $H$ shown in Figure~\ref{fig: hfree 5-con k4 eg} 
  with a $K_4$. 
  The lines  connecting two rectangles indicate that each vertex in one rectangle is adjacent
  to all vertices in the other rectangle. 
  If there is no line shown between two rectangles, then the vertices in them are non-adjacent, with the exceptions -- (i)
  all the vertices in a red rectangle (dashed) together form a clique; (ii) the rectangles in each green rectangle (dashed) are adjacent.}
  \label{fig: hfree 5-con k4}
  \end{figure}

An example of the construction is shown in Figure \ref{fig: hfree 5-con k4} for a graph $H$ given in Figure~\ref{fig: hfree 5-con k4 eg}.
    Keeping a module isomorphic to $\overline{H}$ guarantees that not all vertices in the module is present in a solution $S$ of $G_\Phi$ 
    (i.e., $G_\Phi\oplus S$ is $H$-free). 
    The purpose of variable gadget $X_i$ is to make sure that both $x_i$ and $\overline{x_i}$ are not placed in a solution $S$, 
    so that we can assign \TRUE\ to all literals corresponding to literal vertices placed in $S$, to get a valid truth assignment
    for $\Phi$. On the other hand, any truth
    assignment assigning \TRUE\ to at least two literals per clause makes sure that the set $S$ formed by choosing literal vertices 
    corresponding to \TRUE\ literals destroys copies of $H$ formed by clause gadgets $C_i$ and the corresponding sets $L_i$ of literal vertices.


\begin{lemma}\label{lem 5-conn if}
    Let  $\Phi$ be a yes-instance of \FSAT\  and $\psi$ be a truth assignment satisfying $\Phi$.
    Then $G_{\Phi} \oplus S$ is $H$-free where $S$  is the set of literal vertices whose corresponding literals were assigned \TRUE\ by $\psi$. 
\end{lemma}

\begin{proof}
    Let $G_{\Phi} \oplus S$ contain an $H$ 
    induced by $A$ (say). Since $H$ is a prime graph and $\overline{H}$ is not isomorphic to $H$, 
    $|A\cap Y|\leq 1$ where $Y$ is a module isomorphic to $\overline{H}$.
    Thus, $\mid A\cap X_{ij}\mid $ is at most one.
    Therefore, since $\{x_i, \overline{x_i}\}$ is not a subset of $S$, we obtain that $X_i$ does not have an induced $H$ in $G_\Phi\oplus S$.
    Recall that, the vertices in $X_{ij}$ (for $3\leq j\leq t$) are non-adjacent to $V(G)\setminus X_i$, and $H$ is 5-connected. 
    This implies that $A\cap (X_i\setminus \{x_i, \overline{x_i}\}) = \emptyset$. 
    
    Since $C_i$ contains $t-4$ sets of $\overline{H}$s, $|C_i\cap A|\leq t-4$. 
    Now assume that $A$ contains vertices from two clause gadgets $C_i$ and $C_j$.
    Since the vertices in $C_i$ are only adjacent to the four literal vertices corresponding to the clause $C_i$,
    and $H$ is 5-connected, removing the four literal vertices corresponding to $C_i$ disconnects the graph which is not possible --note that $C_i$ and $C_j$ are non-adjacent. 
    Hence, $A$ contains vertices from at most one clause gadget $C_i$.
    
     Note that $L$ induces a $K_n \times nK_1$ in $G_\Phi\oplus S$, if $V'$ induces a clique, and induces a $K_n+nK_1$ in 
     $G_\Phi\oplus S$, if $V'$ induces an independent set.
     Therefore, $H$ is not an induced subgraph of the graph induced by $L$ in $G_\Phi\oplus S$. 
      Recall that the vertices in $A\cap C$ are from at most one clause gadget $C_i$, and at most one vertex from each of the sets $C_{ij}$ in $C_{i}$ is in  $A\cap C_i$ . 
     We know that $C_i$ is non-adjacent to all literal vertices corresponding to the literals not in the clause $C_i$, and $H$ is 5-connected. 
     Therefore, $A\cap L=\{y_{i,1}, y_{i,2}, y_{i,3}, y_{i,4}\}$.  
     Since at least two vertices in $A\cap L$ is in $S$, the graph induced by $A$ in $G\oplus S$ is not isomorphic to $H$.
\end{proof}

\begin{lemma}\label{lem 5 conn onlyif}
 Let $\Phi$ be an instance of \FSAT. If $G_{\Phi} \oplus S$ is $H$-free  for some $S\subseteq V(G_{\Phi})$, then there exists a truth assignment satisfying $\Phi$.
 \end{lemma}
 
  \begin{proof}
     Let  $G_{\Phi} \oplus S$ be $H$-free for some $S \subseteq V(G_{\Phi})$.  We want to find a satisfying truth assignment of $\Phi$.  
     Since each of the $C_{ij}$s in $C_i$, for $1\leq i\leq m$ and $1\leq j\leq t-4$, induces an $\overline{H}$, 
     there is at least one vertex in each $C_{ij}$ which is not in $S$. 
     Then, if at least two vertices from $L_i$ are not in $S$, then there is an induced $H$ by vertices in $L_i$ and one vertex each from 
     $C_{ij}\setminus S$, for $1\leq j\leq t-4$.
     Therefore, at least two vertices from $L_i$ are in $S$.
     Next we prove that $\{x_i, \overline{x_i}\}$ is not a subset of $S$. 
     For each $X_{ij}$ (for $3\leq j\leq t$), since each of them induces an $\overline{H}$, at least one vertex is not in $S$.
     Then, if both $x_i$ and $\overline{x_i}$ are in $S$, then there is an $H$ induced by $x_i, \overline{x_i}$, and one vertex each from 
     $X_{ij}\setminus S$, for $3\leq j\leq t$.
     Now, it is straight-forward to verify that assigning \TRUE\ to every literal $x_i$ such that $x_i\in S$, is a valid satisfying truth assignment of $\Phi$.
\end{proof}    

Now, Theorem~\ref{thm:5-conn} follows from Lemma~\ref{lem 5-conn if} and Lemma~\ref{lem 5 conn onlyif}.

\section{Trees}
\label{sec:trees}

By $\mathcal{T}$ we denote the set $\mathcal{P} \cup \mathcal{T}_1\cup \mathcal{T}_2\cup \mathcal{T}_3\cup \mathcal{C}$, where 
$\mathcal{P}=\{P_x\mid 1\leq x\leq 5\}$,
$\mathcal{T}_1 = \{K_{1,x}\mid 1\leq x\leq 4\}$, 
$\mathcal{T}_2=\{T_{x,y}\mid 1\leq x\leq y\leq 4\}$, 
$\mathcal{T}_3=\{T_{1,0,1}, T_{1,0,2}\}\cup \{T_{x,y,z}\mid x=1, 1\leq y\leq 4, 1\leq z\leq 5\}$, and
$\mathcal{C} = \{C_{1,1,1}, C_{1,1,2}, C_{1,1,3}, C_{1,2,2}, C_{1,2,3}, C_{1,3,3}, C_{2,2,2}, C_{2,2,3}\}$.
These sets denote the paths, stars, bistars, tristars, and subdivisions of claw not handled by our reductions.

We note that $|\mathcal{P}| = 5$, $|\mathcal{T}_1| = 4$, $|\mathcal{T}_2| = 10$, $|\mathcal{T}_3| = 22$, and $|\mathcal{C}| = 8$.
However, a star graph $K_{1,x}$ is a path in $\mathcal{P}$ if $x\leq 2$, 
the bistar graph $T_{1,1}$ is the path $P_4$, 
the tristar graphs $T_{1,0,1}$ is the path $P_5$, and
the subdivision of claw $C_{1,1,1}$ is the star graph $K_{1,3}$,
$C_{1,1,2}$ is the bistar graph $T_{1,2}$,
$C_{1,1,3}$ is the tristar graph $T_{1,0,2}$, and
$C_{1,2,2}$ is the tristar graph $T_{1,1,1}$. Therefore, $|\mathcal{T}| = 41$,
and the tree of maximum order in $\mathcal{T}$ is $T_{1,4,5}$ with 13 vertices.
We prove the following theorem in this section.

\begin{theorem}
\label{thm:tree}
    Let $T$ be a tree not in $\mathcal{T}$. Then \SCTF{T}\ is NP-Complete. \TETHS.
\end{theorem}
This task is achieved in seven sections. In the first section, we prove that there is a linear reduction from \SCTF{T'} to \SCTF{T}, where $T$
is a prime tree and $T'$ is its internal tree. In the second section, we deal with trees with at least 4 leaves and at least 3 internal vertices,
and satisfying
some additional constraints. Then in the third and the fourth sections, we prove the hardness for bistars and tristars respectively, leaving behind a finite number 
of open cases. The fifth section proves the hardness for $P_6$, thereby leaving only one unsolved case ($P_5$) among paths. The sixth section settles subdivions of claw except for a finite number of cases. We combine all these results in seventh section to prove Theorem~\ref{thm:tree}.
\subsection{Removing leaves}
\label{sub:leaves added}
\begin{figure}[!htbp]
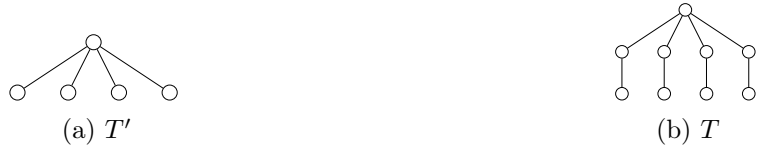

    \centering
     \begin{subfigure}[b]{0.45\textwidth}
          \centering
          \resizebox{0.3\linewidth}{!}{\input{figs/Tree/eg_leaves_added_T1}}  
          \caption{$T'$}
          \label{fig:T'added}
     \end{subfigure}
     \begin{subfigure}[b]{0.45\textwidth}
          \centering
          \resizebox{0.25\linewidth}{!}{\input{figs/Tree/eg_leaves_added}}  
          \caption{$T$}
          \label{fig:Tadded}
     \end{subfigure}
     \caption{An example of $T'$ and $T$}
     \label{fig: eg leaves added}
\end{figure}

In this section, with a very simple reduction, we prove that the hardness transfers from $T'$ to $T$,
where $T$ is a prime tree and $T'$ is its internal tree.
We use Construction~\ref{cons tree leaves added} for the reduction.
See Figure~\ref{fig: eg leaves added}, for an example of $T$ and $T'$.

\begin{lemma}
   \label{lem:leaf-removal}
   Let $T$ be a prime tree and let $T'$ be its internal tree. Then there is a linear reduction from \SCTF{T'}\ to \SCTF{T}.
\end{lemma}

\begin{figure}[ht]
\centering
 \input{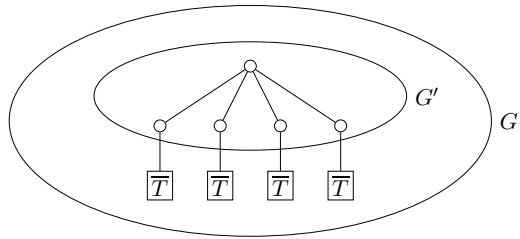}
  \caption{An example of Construction \ref{cons tree leaves added}}
    
    \label{fig tree leaves added}
  \end{figure}%
  
\begin{construction}
\label{cons tree leaves added}
    Let $(G',T)$ be the input to the construction, where $G'$ is a graph and $T$ is a prime tree.
    The graph $G$ is constructed from $G'$ as follows:
    for every vertex $u$ of $G'$, introduce a $\overline{T}$, denoted by $W_u$ in the neighbourhood of $u$
    (see Figure~\ref{fig tree leaves added} for an example).
\end{construction}

\begin{lemma}
\label{lem leaves addeed if}
    If $G' \oplus S' \in \mathcal{F}(T')$ for some $S' \subseteq V(G')$, then   $G \oplus S' \in \mathcal{F}(T)$.
\end{lemma}

\begin{proof}
     Let a $T$ be induced by a set $A$ in $G\oplus S'$.
     Note that $T$ is a prime graph and $\overline{T}$ is not isomorphic to $T$.
     Thus, $W_u$ does not induce a $T$.
     For any vertex $v\in W_u$, the only neighbor of $v$ in  $V(G)\setminus W_u$ is $u$. 
     Hence $A\cap V(G')$ is nonempty. 
     Let $u\in A\cap V(G')$. 
     Recall that the vertices in $W_u$ are the only neighbors of $u$ in $V(G)\setminus V(G')$.
     Since $T$ is a prime tree and $W_u$ induces a module in $G$, $|A\cap W_u|\leq 1$. 
     Thus, $A\cap W_u$ cannot contain any internal vertex of $T$ which implies that $G'\oplus S'$ contains a $T'$.
     However, that is not possible according to the statement of the lemma. 
\end{proof}

\begin{lemma}
\label{lem leaves added onlyif}
   If $G \oplus S \in \mathcal{F}(T)$ for some $S \subseteq V(G)$, then   $G' \oplus S' \in \mathcal{F}(T')$, where $S' = S \cap V(G')$.
\end{lemma}

\begin{proof}
   If $G' \oplus S'$ contains a $T'$ induced by 
  $A$ (say), then $G \oplus S$ will contain  a $T$ unless for at least one  vertex $u\in A$, all vertices of $W_u$ belong to $S$. 
  However, in that case we will have a $T$ induced by $W_u$ in $G\oplus S$, which is a contradiction.
\end{proof}

Now, Lemma~\ref{lem:leaf-removal} follows from Lemma~\ref{lem leaves addeed if} and Lemma~\ref{lem leaves added onlyif}.

\subsection{Trees with at least 4 leaves and 3 internal vertices}
\label{sec: 4 leaves 3 internal}

In this section, we prove hardness results for \SCTF{T}, when $T$ is a tree with at least 4 leaves and at least 3 
internal vertices, and satisfying some additional constraints.
The reduction is from \KSAT.

\begin{theorem}
    \label{thm:i4l4}
    Let $T$ be a tree with at least 4 leaves and at least 3 internal vertices. Let $T'$ be the internal tree of $T$. 
    Assume that the following properties are satisfied.
    \begin{enumerate}[label=(\roman*)]
        \item If $T'$ is a star graph, then at least one of the following conditions are satisfied:
        \begin{enumerate}
            \item every leaf of $T'$ has at least two leaves of $T$ as neighbors, or
            \item the center of the star $T'$ has no leaf of $T$ as neighbor, or
            \item $T$ is either a $C_{1,2,2,2,}$ or a $C_{1,2,2,2,2}$.
        \end{enumerate}
        \item There are no two adjacent vertices of degree 2 in $T$ such that neither of them is adjacent to any leaf of $T$.
    \end{enumerate}
    Then \SCTF{T}\ is NP-Complete. \TETHS.
\end{theorem}

Let $T$ be a tree and $T'$ be its internal tree. Assume that $T$ satisfies the conditions of Theorem~\ref{thm:i4l4}.  
Let $T$ has $p$ internal vertices and $k$ leaves. Then, $T$ has $t=p+k$ vertices, and $T'$ has $p$ vertices.
Let $V(T)=\{v_1,v_2,\ldots, v_t\}$, where $\{v_1,v_2,\ldots,v_p\}$ forms the internal vertices. 
Without loss of generality, we assume that $v_1$ and $v_2$ are adjacent.
Let $\mathcal{M}=\{M_1,M_2,\ldots,M_{t'}\}$ be the modular decomposition of $T$, where $t'=k'+p$, where $k'$ is the number of modules containing leaves.
Let $\{M_1,M_2,\ldots, M_p\}$ forms the modules ($M_i= \{v_i\}$, for $1\leq i\leq p$) containing the internal vertices, and $\{M_{p+1}, M_{p+2}, \ldots, M_{t'}\}$
forms the modules containing leaves.
Let $r$ be the maximum size of modules containing leaves, i.e., $r= max_{i={p+1}}^{i=t'}\{|M_i|\}$.
Let $Q_T$ be the quotient tree of $T$. By $R$, we denote the graph $\overline{Q_T}[rK_r]$, i.e., 
$R$ is the graph obtained from $\overline{Q_T}$ by replacing each vertex by a disjoint union of $r$ copies of $K_r$.

We use Construction~\ref{cons 4 leaves and 4 internal} for the reduction from \KSAT. 
The reduction is very similar to the reduction used to handle 5-connected graphs in Section~\ref{sub:5-conn}.

\begin{figure}[!htbp]
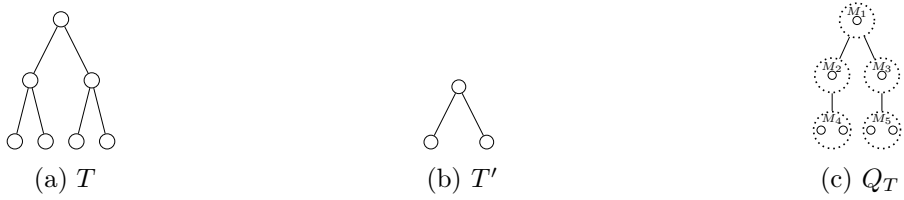

    \centering
    \begin{subfigure}[b]{0.3\textwidth}
          \centering
          \resizebox{0.3\linewidth}{!}{\input{figs/Tree/eg1_4_leaves_3_internal}}  
          \caption{$T$}
          \label{fig:t}
     \end{subfigure}
    \begin{subfigure}[b]{0.3\textwidth}
          \centering
          \resizebox{0.2\linewidth}{!}{\input{figs/Tree/eg2_4leaves_3_internal}}  
          \caption{$T'$}
          \label{fig:t'}
     \end{subfigure}
     \begin{subfigure}[b]{0.3\textwidth}
          \centering
          \resizebox{0.25\linewidth}{!}{\input{figs/Tree/eg_4_leaves_3_internal}}  
          \caption{$Q_T$}
          \label{fig:qt}
     \end{subfigure}
     
     \caption{An example of a Tree $T$ (\ref{fig:t}) which satisfies the properties of Theorem~\ref{thm:i4l4}, its internal tree $T'$ (\ref{fig:t'}), and its quotient graph 
     $Q_T$ (\ref{fig:qt}) in which five modules of $T$ are represented with circles (dotted). 
     The bold circles represent the vertices in $T$.
     The lines connecting two circles (dotted) indicate that each vertex in one circle is adjacent to all vertices in the other circle.}
    \label{fig: eg 4 leaves 4 internal}
  \end{figure}
\begin{figure}[!htbp]
  \centering
    \input{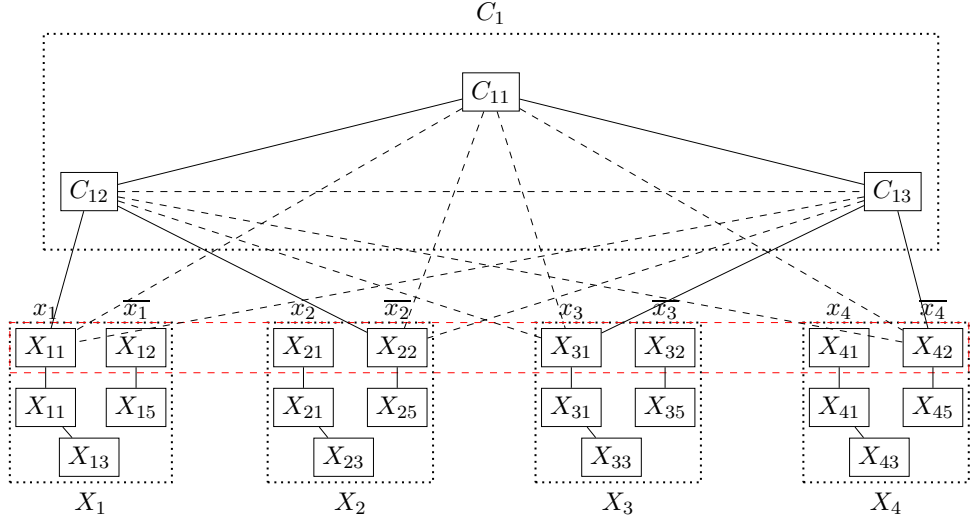}
    \caption{An example of Construction \ref{cons 4 leaves and 4 internal} for the formula 
    $\Phi = C_1$  where $C_1=x_1\lor \overline{x}_2\lor x_3\lor \overline{x}_4$ corresponding to the tree $T$ shown in Figure~\ref{fig: eg 4 leaves 4 internal}.
    The bold lines (respectively dashed lines) connecting two rectangles indicate that 
     each vertex in one rectangle is adjacent (respectively non-adjacent) to all vertices in the other rectangle. 
     If there is no line shown between two rectangles, then the vertices in them are adjacent, 
     with an exception --  all the vertices in the red rectangle (dashed) together form an independent set. 
     Similarly,  if there is no line shown between two rectangles in the dotted rectangles, then the rectangles in them are non-adjacent.}
    \label{fig 4 leaves 4 internal}
\end{figure}

%
\begin{construction}
\label{cons 4 leaves and 4 internal}
    Let $\Phi$ be a \KSATO\ formula with $n$ variables $X_1, X_2, \cdots, X_n$, and $m$ clauses $C_1, C_2, \cdots,$ $C_m$. 
    We construct the graph $G_{\Phi}$ as follows.
    
 \begin{itemize}
     \item  For each variable $X_i$ in $\Phi$, the variable gadget, 
     also named  $X_i$, consists of two special sets $X_{i1}=\{x_i\}, X_{i2}=\{\overline{x_i}\}$, and 
     $t'-2$ other sets $X_{i3}, X_{i4},\ldots, X_{it'}$, where each of the set in $\{X_{i3}, X_{i4},\ldots, X_{it'}\}$ induces an $R$.
     We have $X_i=\bigcup_{j=1}^{j=t'}X_{ij}$.
     The sets $X_{ia}$ and $X_{ib}$ are adjacent if and only if $M_a$ and $M_b$ are adjacent. 
     We remove the edge between $X_{i1}$ and $X_{i2}$  to end the 
     construction of the variable gadget (recall that $v_1$ and $v_2$ are 
     adjacent in $T$).
     Let $X=\bigcup_{i=1}^{i=n}X_i$. 
     The vertices $x_i$ and $\overline{x_i}$ are called literal vertices, and $L$ is the set of all literal vertices.  
     The set $L$ forms an independent set of size $2n$.

     \item For each clause $C_i$ in $\Phi$ of the form  $(\ell_{i1}\lor \ell_{i2}\lor \ell_{i3}\lor \ldots \lor \ell_{ik})$, 
     the clause gadget, also named $C_i$, consists of $p$ copies of $\overline{T}$s called  $C_{i1}$, $C_{i2}, \ldots C_{ip}$. 
     The set of union of all clause gadgets is denoted by $C$. Let the $k$ vertices introduced (in the previous step) 
     for the literals $\ell_{i1}, \ell_{i2}, \ell_{i3}, \ldots \ell_{ik}$ be denoted by  
     $L_i=\{y_{i1},y_{i2},y_{i3}, \ldots y_{ik}\}$. 
     Make the adjacency among these sets $C_{ij}$s and the corresponding literal vertices 
     in $L_i$ in such a way that, taking one vertex from each set $C_{ij}$ along with the literal vertices 
     $L_i$ induces a $T$, where the vertices in $L_i$ correspond to the $k$ leaves of $T$.
     We observe that $C_i$ is obtained from the internal tree 
     of $T$ where each vertex is replaced by $\overline{T}$.
     In addition to this, every vertex in $C_i$ is adjacent to all literal vertices corresponding to literals not in $C_i$.

     \item For all $i \neq j$, the set $C_i$ is adjacent to the set $C_j$.
     
     \item The vertices in $X_{i}\setminus \{x_i,\overline{x_i}\}$ are adjacent to $V(G)\setminus X_i$, for $1\leq i\leq n$.  
     
     This completes the construction of the graph $G_{\Phi}$ (see Figure~\ref{fig 4 leaves 4 internal} for an example). 
  
 \end{itemize}
    
\end{construction}

Observation~\ref{obs:R} says that $T$ cannot be an induced subgraph of any of the $X_{ij}$s and any of the $C_i$s, and that any solution of $G_\Phi$
leaves an independent set of size $r$ untouched in $X_{ij}$, which induces an $R$.
\begin{observation}
\label{obs:R}
\begin{enumerate}[label=(\roman*)]
    \item\label{obs:R:i} $T$ is not an induced subgraph of $R$.
    \item\label{obs:R:ii} $T$ is not an induced subgraph of $C_i$ for any $1\leq i\leq m$.
    \item\label{obs:R:iii} Let $S$ be a subset of vertices of $G_\Phi$ such that $G_\Phi\oplus S$ is $T$-free.
    Then there is an independent set of size $r$ in $X_{ij}\setminus S$ for any $1\leq i\leq n$ and $3\leq j\leq t'$.
\end{enumerate}
\end{observation}
\begin{proof}
    To prove \ref{obs:R:i} by contradiction, assume that $A$ induces a $T$ in $\overline{Q_T}[rK_r]$. 
     Clearly, there is no induced copy of $T$ in $rK_r$. Therefore, $A$ must have nonempty intersection with more than one copy of $rK_r$. 
     Since $Q_T$ and $\overline{Q_T}[rK_r]$ have $t'$ modules each, $A$ must have nonempty intersection with every copy of $rK_r$. 
     Then the quotient graph of the graph 
     induced by $A$ is $\overline{Q_T}$, which is not isomorphic to $Q_T$, as no nontrivial tree is self-complementary. 
     Therefore, $A$ does not induce $T$.

     To prove \ref{obs:R:ii} by contradiction, assume that $A$ induces a $T$ in $C_i$.
     We recall that $C_i$ represents the internal tree of $T$, where
     each vertex is replaced by $\overline{T}$. Since no nontrivial tree is self-complementary, $A$ must have nonempty
     intersection with at least two sets $C_{ij}$ and $C_{i\ell}$. Then, the graph induced by $A$ has only at most $p$
     modules, which contradicts with the fact that $T$ has $t'>p$ modules.
     
     To prove \ref{obs:R:iii} by contradiction, assume that there is no independent set of size $r$ in $X_{ij}$ untouched by $S$.
     This implies that, for every copy of $rK_r$ in $X_{ij}$, one clique of size $r$ is in $S$. Let $A$ be a union of such cliques, one 
     from each copy of $rK_r$. The set $A$ induces $\overline{Q_T}[K_r]$ in $G_\Phi$, and $Q_T[rK_1]$ in $G_\Phi\oplus S$. 
     Since $T$ is an induced subgraph of $Q_T[rK_1]$, we obtain a contradiction.
\end{proof}

Before proving the forward direction of the correctness of the reduction, we separately handle three cases in the forward direction using Lemmas~\ref{lem 4 leaves 3 internal if - special - i}, \ref{lem 4 leaves 3 internal if - special - ii}, and \ref{lem 4 leaves 3 internal if - special - iii}.

\begin{lemma}
    \label{lem 4 leaves 3 internal if - special - i}
   Let $\Phi$ be a yes-instance of \KSAT\  and $\psi$ be a truth assignment satisfying $\Phi$. 
   Let $S$ be the set of literal vertices whose corresponding literals were assigned \TRUE\ by $\psi$.
   Then there exists no set $A$ such that $A$ induces a $T$ in $G_\Phi\oplus S$, and   $A\subseteq C\cup L$ and $|A\setminus (C_i\cup L_i)| = 1$,
   and $|A\cap C_i|\geq 2$ (for some $1\leq i\leq n$). 
\end{lemma}
\begin{proof}
     Assume for a contradiction that there exists such a set $A$. 
     Let $A\setminus (C_i\cap L_i) = \{w\}$.
    Clearly, $A\cap C_i$ is an independent set - otherwise, there is a triangle formed by $w$ and two adjacent vertices in $A\cap C_i$.
    Further, $w$ is an internal vertex of the tree induced by $A$, as $w$ is adjacent to every vertex in $C_i$ and $|C_i\cap A|\geq 2$.
    We recall that the independent number of a tree $T$ is at most $|V(T)|-1$, which is achieved when the tree is a star graph.   
    Since $C_i$ corresponds to the internal tree $T'$ (having $p$ vertices) of $T$, 
    we obtain that $A\cap C_i$ can have vertices from only $p-1$ modules, say $C_{i1}, C_{i2}, \ldots, C_{i{p-1}}$ 
    of $C_i$, and $A$ can have nonempty intersection with $p-1$ sets in $C_i$ only when $T'$ is a star graph. 
    Since $A\cap L_i$ induces a subgraph of $K_2 + (k-2)K_1$ (at least two vertices of $L_i$ are in $S$), 
    all the leaves of the tree induced by $A$ cannot be from $L_i$. Therefore, at least one leaf is from $C_i$. 
    Since $L_i$ can contribute only one internal vertex, $p-2$ internal vertices of the tree must be from $C_i$ (the remaining internal 
    vertex is $w$). This implies that $A$    has 
    nonempty intersection with $p-1$ sets say $C_{i1}, C_{i2}, \ldots, C_{i{p-1}}$ 
    (a leaf and an internal vertex cannot come from a set $C_{i\ell}$, which is a module). 
    Hence the internal tree $T'$ of $T$ is a star graph.
    Let $x, x'\in A\cap L_i$ be such that $x$ is the internal vertex which is adjacent to $x'$. 
    The vertex $x$ cannot get a leaf from $C_i$ as $w$ is adjacent to every vertex in $C_i$.
    Then $x$ is an internal vertex of the tree having a single leaf ($x'$) of $T$ as neighbor. 
    Further, the center vertex ($w$) of the internal star has a leaf of $T$ (from $C_i$) as neighbor.
    Then, by the assumption in the statement of Theorem~\ref{thm:i4l4}, $T$ is either $C_{1,2,2,2}$ or $C_{1,2,2,2,2}$.
    Let $T$ be $C_{1,2,2,2}$. 
    Let $C_{i1}$ correspond to the root of the internal tree and $C_{i2}, C_{i3}$, and $C_{i4}$
    correspond to the leaves of the internal tree $T'$ of $T$. Let $y_{i1}, y_{i2}, y_{i3}$, and $y_{i4}$
    correspond to the leaves of $T$ adjacent to $C_{i1}, C_{i2}, C_{i3}$, and $C_{i4}$ respectively.
    Clearly, $A$ contains $c_{i2}\in C_{i2}, c_{i3}\in C_{i3}, c_{i4}\in C_{i4}$, and all vertices in $L_i=\{y_{i1}, y_{i2}, y_{i3}, y_{i4}\}$.
    Then, all vertices in $A\cap C_i$ are internal vertices of the tree, which is a contradiction.
    The case when $T$ is $C_{1,2,2,2,2}$ can be handled in a similar way.
\end{proof}

\begin{lemma}
    \label{lem 4 leaves 3 internal if - special - ii}
   Let $\Phi$ be a yes-instance of \KSAT\  and $\psi$ be a truth assignment satisfying $\Phi$. 
   Let $S$ be the set of literal vertices whose corresponding literals were assigned \TRUE\ by $\psi$.
   Then there exists no set $A$ such that $A$ induces a $T$ in $G_\Phi\oplus S$, and  $A\subseteq C_i\cup C_j\cup L$, and 
   $|A\cap C_i| = |A\cap C_j| = 1$ (for some $1\leq i\neq j\leq n$). 
\end{lemma}
\begin{proof}
    Assume that $A\cap C_i = \{c_i\}$ and $A\cap C_j = \{c_j\}$.
    Since the rest of the vertices in $A$ are from $L$, there is at most one internal vertex from $L\cap A$. If there are no 
    internal vertices from $A\cap L$, then $T$ has only at most two internal vertices, a contradiction.
    Therefore, there is exactly one internal vertex from $L\cap A$. Then $A$ induces a tristar graph. Without loss of generatlity,
    assume that $c_i$ is the center of the internal $P_3$ and $x$ is the internal vertex from $L$, and $x'\in L$ be the leaf adjacent to $x$.
    Assume that $c_{i}$ has no attached leaf, i.e., $T$ is the tristar graph $T_{1,0,k-1}$. 
    Since none of the leaves are adjacent to $c_{i}$, 
    all $k$ leaves are from $L_i$, i.e., $L_i\subseteq A$ (recall that $c_{i}$ is adjacent to all literal vertices correspond to literals 
    not in $C_i$). This is a contradiction, as there is an edge induced by $L_i$ in $G_\Phi\oplus S$.
    Therefore, $c_{i}$ has some attached leaves in the tree induced by $A$. Hence, by the condition (i) of Theorem~\ref{thm:i4l4}, 
    $T$ is either $C_{1,2,2,2}$ or $C_{1,2,2,2,2}$. These cases give contradictions as then there are more than three internal vertices.
\end{proof}

\begin{lemma}\label{lem 4 leaves 3 internal if - special - iii}
   Let $\Phi$ be a yes-instance of \KSAT\  and $\psi$ be a truth assignment satisfying $\Phi$. 
   Let $S$ be the set of literal vertices whose corresponding literals were assigned \TRUE\ by $\psi$.
   Then $T$ is not an induced subgraph of the graph induced by $C_i\cup L_i$ in $G_\Phi\oplus S$, for any $1\leq i\leq n$. 
\end{lemma}
\begin{proof}
    Assume that $A\subseteq C_i\cup L_i$ induces a $T$ in $G_\Phi\oplus S$.
    By Observation~\ref{obs:R}\ref{obs:R:ii}, $A$ is not a subset of $C_i$.
    Clearly, $A\cap L_i$ can have at most one edge, as $L$  
    induces $K_n+nK_1$ in $G_\Phi\oplus S$. 
    No other vertex in $A\cap L_i$
    other than the end vertices of this edge can be an internal vertex of the $T$ induced by $A$ (by construction, 
    no vertex in $L_i$ has two modules $C_{ij}$ and $C_{i\ell}$
    as neighbors as the vertices in $L_i$ correspond to the leaves of $T$).
    Therefore, at least $p-2$ internal vertices are from $A\cap C_i$.
    
    Since $L_i$ cannot contribute all leaves (at least two vertices in $L_i$ are in $S$),
    at least one leaf must be from $C_i$. Therefore, $C_i$ contributes only at most $p-1$ internal vertices.
    Therefore, there are two vertices $u,v\in L_i\cap A$ such that $uv$ is an edge in $G_\Phi\oplus S$.
    Assume that $C_i$ contributes exactly $p-1$ internal vertices. 
    Then $A$ has nonempty intersection with all modules $C_{ij}$ in $C_i$. 
    Then, the edge $uv$ in $A\cap L_i$ along with 
    the path through $A\cap C_i$ from the neighbor of $u$ in $A\cap C_i$ and the neighbor of $v$ in $A\cap C_i$
    forms a cycle, which is a contradiction.
    
    Therefore, exactly $p-2$ internal vertices are from $C_i$. Then, both $u$ and $v$ must be internal vertices. 
    Then, only at most $k-2$ leaves are from $A\cap L_i$. 
    Let $C_{ia}$ and $C_{ib}$ be the two modules in $C_i$ which do not contribute internal vertices. 
    Each of $C_{ia}$ and $C_{ib}$ contributes only at most two leaf vertices ($\overline{T}$ is $3K_1$-free).
    If both $C_{ia}$ and $C_{ib}$ contribute leaves, then there is a cycle 
    as described in the previous case. 
    Therefore, exactly $k-2$ leaves are from $A\cap L_i$ and two leaves are from one module, say $C_{ia}$.
    Let $w_1$ and $w_2$ be the leaves contributed by $C_{ia}$.
    Assume that $v$ is adjacent to  $C_{ia}$. 
    Let $u'$ be the neighbor of $u$, other than $v$, in the tree. Let $u'\in C_{i\ell}$.
    We note that $a\neq \ell$ (otherwise, there is a triangle formed by $u,v$, and a vertex in $A\cap C_{ia}$).
    Let $T''$ be the tree induced by $A$. By \textit{leaf-degree} of a vertex in a tree, we mean the number of 
    leaves adjacent to that vertex in the tree. The \textit{leaf-degree sequence} of a tree is the 
    non-decreasing sequence of leaf-degrees of vertices of the tree.
    We claim that there is a mismatch in the leaf-degree sequences of $T$ and $T''$, which provides a contradiction.
    Let $U$ be the set containing one vertex each from the modules of $C_i$, except from $C_{ia}, C_{ib}$, and $C_{i\ell}$.
    We know that the set $U$,
    two vertices $w_1, w_2\in C_{ia}$, $u'\in C_{i\ell}$, and the vertices in $L_i$
    induce $T''$ in $G_\Phi\oplus S$.
    Further, $L_i\cup U\cup \{u', w_1, c_{ib}\}$ induces a $T$ in $G_\Phi$, where $c_{ib}$ is any vertex in $C_{ib}$.
    Every vertex in $U\cup (L_i\setminus \{u,v\})$ has the same leaf-degree in $T$ and $T''$.
    The leaf-degree of $u'$ is one less in $T''$ than that in $T$ ($u$ is not a leaf in $T''$).
    The leaf-degree of $u$ is 0 in both $T$ and $T''$. 
    The leaf-degree of $v$ is 0 in $T$ and 2 in $T''$ ($v$ is a leaf in $T$, and is adjacent to 2 leaves - $w_1$ and $w_2$ - in $T''$).
    The leaf-degree of $c_{ia}$ is 1 in $T$ ($v$ is the only leaf, otherwise there will be a $C_4$ in $T''$ 
    induced by $w_1,w_2,v$, and the other leaf) and $w_1$ and $w_2$ have leaf-degree 0 in $T''$.
    The leaf-degree of $c_{ib}$ is 0 in $T$ (if it is adjacent to some leaf, then that leaf in $L_i$ has no neighbor in $T''$, which is not true).
    This implies that the leaf-degree sequences of $T$ and $T''$ are not the same (see Figure~\ref{fig:uv-leafdegree1} for an example).
    
\begin{figure}[!htbp]
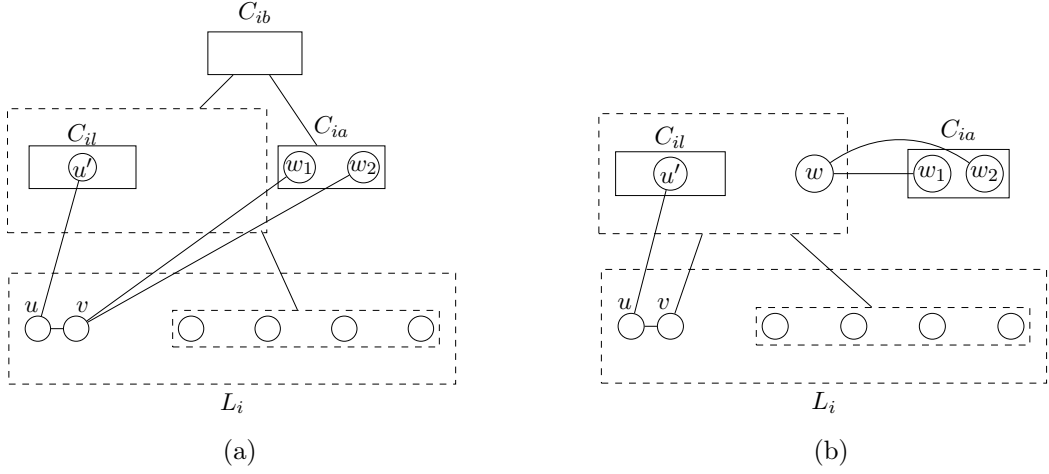

    \centering
    \begin{subfigure}[b]{0.45\textwidth}
          \centering
          \resizebox{0.8\linewidth}{!}{\input{figs/Tree/uv-leafdegree}}  
          \caption{}
          \label{fig:uv-leafdegree1}
     \end{subfigure}
    \begin{subfigure}[b]{0.45\textwidth}
          \centering
          \resizebox{0.8\linewidth}{!}{\input{figs/Tree/uv-degree2}}  
          \caption{}
          \label{fig:uv-degree2}
     \end{subfigure}
    
     \caption{The cases discussed in Lemma~\ref{lem 4 leaves 3 internal if - special - iii}, when $C_i$ contributes exactly two leaves.}
    \label{fig:uv-leafdegree}
\end{figure}
    
    Therefore, there is a vertex $w\in A\cap C_i$ which is adjacent to the two leaf vertices from $C_{ia}$ (see Figure~\ref{fig:uv-degree2} for an example).
    Then $u$ and $v$ form two adjacent internal vertices with degree 2 such that neither $u$ nor $v$ is adjacent to a leaf of $T$, 
    which contradicts with condition (ii) of Theorem~\ref{thm:i4l4}.
\end{proof}

With Lemma~\ref{lem 4 leaves 3 internal if - special - i}, Lemma~\ref{lem 4 leaves 3 internal if - special - ii}, and 
Lemma~\ref{lem 4 leaves 3 internal if - special - iii}, we are ready to prove the forward direction of the reduction.
\begin{lemma}\label{lem 4 leaves 4 internal if}
   Let $\Phi$ be a yes-instance of \KSAT\  and $\psi$ be a truth assignment satisfying $\Phi$. 
   Then $G_{\Phi} \oplus S\in \mathcal{F}( T)$ where $S$  is the set of literal vertices whose corresponding literals were assigned \TRUE\ by $\psi$. 
\end{lemma}

\begin{proof}
    Let $G_{\Phi} \oplus S$ contain a $T$ induced by $A$ (say).
    We prove the lemma with the help of a set of claims.
    
    Claim 1: $A$ is not a subset of $X_i$, for $1\leq i\leq n$.
    
    Assume that $A$ is a subset of $X_i$.
    By Observation~\ref{obs:R}\ref{obs:R:i}, $A$ is not a subset of $X_{ij}$, for $1\leq j\leq t'$.
    Therefore, $A$ has nonempty intersection with at least two sets $X_{ij}$ and $X_{i\ell}$.
    Since $X_{i}$ induces a graph with at most $t'$ modules, and $T$ has $t'$ modules, 
    $A$ has nonempty intersection with all sets $X_{ij}$ ($1\leq j\leq t'$). 
    Since $\{x_i, \overline{x_i}\}$ is not a subset of $S$, we obtain that the quotient graph of the graph induced by $A$, which is a forest of two trees, 
    is not isomorphic to $Q_T$, which is a contradiction.

    Claim 2: Let $X_i'= X_i\setminus \{x_i, \overline{x_i}\}$ and $\overline{X_i} = V(G_\Phi)\setminus X_i$.
    If $|A\cap X_i'|\geq 1$, then $A\cap \overline{X_i}=\emptyset$. Similarly, if $|A\cap \overline{X_i}|\geq 1$,
    then $A\cap X_i'=\emptyset$. 
    
    For a contradiction, assume that $A$ contains at least one vertex from $X_i'$ and at least one vertex from $\overline{X_i}$.
    Since $X_i'$ and $\overline{X_i}$ are adjacent, either $|A\cap X_i'|=1$ or $|A\cap \overline{X_i}|=1$.
    
    Assume that $A\cap X_i' = \{u\}$. Since $T$ has at least 3 internal vertices and $(A\cap X_i')\cup (A\cap \overline{X_i})$ induces a star graph,
    both $x_i$ and $\overline{x_i}$ are in $A$. 
    Then $T$ is the tristar graph $T_{1,t-5,1}$, which is a contradiction as condition (i) of Theorem~\ref{thm:i4l4}  is not satisfied. 
    Assume that $A\cap \overline{X_i} = \{u\}$. Then with the same argument as given above, we obtain that the 
    graph induced by $A$ is $T_{1,t-5,1}$, which is a contradiction.
    
    Claim 3: $A$ is not a subset of $L$, the set of all literal vertices.
    
    This follows from the fact that $L$ induces a $K_n+nK_1$ in $G_\Phi\oplus S$.
    
    Claim 4: $A$ cannot have nonempty intersections with three distinct clause gadgets $C_i$, $C_j$, and $C_\ell$.

    Claim 5: There exists no $C_i$ and $C_j$ ($i\neq j$) such that $|A\cap C_i|\geq 2$ and $|A\cap C_j|\geq 2$.
    
    Claim 4 and 5 follow from the fact that $C_i$ and $C_j$ are adjacent for $i\neq j$ and $T$ does have neither a triangle nor a $C_4$.

    Claim 6: $A$ is not a subset of $C$.
    
    For a contradiction, assume that $A\subseteq C$.
    By Claim 4, $A$ cannot have nonempty intersections with three distinct clause gadgets $C_i$, $C_j$, and $C_\ell$.
    By Observation~\ref{obs:R}\ref{obs:R:ii}, $A$ cannot be a subset of $C_i$. 
    Therefore, $A$ has nonempty intersection with exactly two clause gadgets $C_i$ and $C_j$ in $C$.
    Then $A$ induces a 
    star graph, which is a contradiction as $T$ has at least 3 internal vertices.
    
    Claim 7: If $|A\cap C_i| \geq 2$ and $A\cap C_j\neq \emptyset$ ($i\neq j$), then $(A\cap L)\subseteq L_i$.
    
    Let $u$ be a vertex in $A\cap L\setminus L_i$. Then there is a $C_4$ formed by $u$ and two vertices in $A\cap C_i$ and one vertex from $A\cap C_j$.

    We are ready to prove the lemma. 
    By Claim 1, $A$ is not a subset of $X_i$. 
    By Claim 2, $A$ cannot have vertices from both $X_i\setminus \{x_i, \overline{x_i}\}$ and $\overline{X_i}$.
    This implies that $A\subseteq L\cup C$. 
    By Claim 3, $A$ cannot be a subset of $L$ and by Claim 6, $A$ cannot be a subset of $C$.
    Therefore, $A$ contains vertices from both $L$ and $C$.
    By Claim 4, $A$ cannot have nonempty intersections with three distinct clause gadgets $C_i$, $C_j$ and $C_\ell$.
    Therefore, $A\cap C\subseteq (C_i\cup C_j)$.
    Assume that $A$ has nonempty intersection with both $C_i$ and $C_j$.
    By Claim 5, we can assume that $|A\cap C_j| = 1$ and $|A\cap C_i|\geq 1$.
    Assume that $|A\cap C_i|\geq 2$. Then by Claim 7, $(A\cap L)\subseteq L_i$.
    Then by Lemma~\ref{lem 4 leaves 3 internal if - special - i}, $A$ cannot induce a $T$.
    Let $|A\cap C_i| = |A\cap C_j|=1$. Then by Lemma~\ref{lem 4 leaves 3 internal if - special - ii}, $A$ cannot induce a $T$.
    
    Assume that $A\cap C\subseteq C_i$ for some clause gadget $C_i$.
    Assume that $A\cap C$ has exactly one vertex. Then the rest of the vertices in $A$
    are from $L$ and only one from $L$ can be an internal vertex. Therefore, $T$ has only at most two internal
    vertices, a contradiction. Therefore, $A\cap C_i$ has at least two vertices. 
    If there are at least two vertices in $A\setminus L_i$, then those two vertices along with two vertices in $A\cap C_i$
    forms a $C_4$. Therefore, $A\cap (L\setminus L_i)$ has at most one vertex. 
    Assume that $|A\cap (L\setminus L_i)| = 1$. Then, by Lemma~\ref{lem 4 leaves 3 internal if - special - i}, $A$ cannot induce a $T$.
    Assume that $A\subseteq C_i\cup L_i$. Then we get a contradiction by Lemma~\ref{lem 4 leaves 3 internal if - special - iii}.
\end{proof}

The backward direction of the proof of correctness of the reduction is easy.
\begin{lemma}\label{lem 4 leaves 4 internal onlyif}
    Let $\Phi$ be an instance of \KSAT. If $G_{\Phi} \oplus S\in \mathcal{F}( T)$ for some $S\subseteq V(G_{\Phi})$, then there exists a truth assignment satisfying $\Phi$.
\end{lemma}

\begin{proof}
     Let  $G_{\Phi} \oplus S\in \mathcal{F}( T)$ for some $S \subseteq V(G_{\Phi})$. 
     We want to find a satisfying truth assignment of $\Phi$. 
     We know that each of the sets $C_{ij}$, for $1\leq i\leq m$ and  $1\leq j\leq p$, induces a $\overline{T}$. 
     Therefore, each such set has at least one vertex not in $S$. Hence at least two vertices in $L_i$ must belong to $S$,
     otherwise there is an induced $T$ by vertices in $L_i$ and one vertex each from $C_{ij}\setminus S$, for $1\leq j\leq p$.
     
     Similarly, each set $X_{ij}$, for $1\leq i\leq n$ and $3\leq j\leq t'$, induces a $\overline{Q_T}[rK_r]$.
     By Observation~\ref{obs:R}\ref{obs:R:iii}, there is an independent set of size $r$ untouched by $S$ in $X_{ij}$. 
      Assume that both $x_i$ and $\overline{x_i}$
      are in $S$. Then there is a copy of $Q_T[rK_1]$ in $G_\Phi\oplus S$,
      induced by $\{x_i,\overline{x_i}\}$ and one copy of $rK_1$ from each $X_{ij}\setminus S$ (for $3\leq j\leq t'$).
      Since $T$ is an induced subgraph of $Q_T[rK_1]$, we get a contradiction.
      Therefore, $\{x_i, \overline{x_i}\}$ is not a subset of $S$. 
      Now, it is straight-forward to verify that assigning \TRUE\ to each literal corresponding to the literal vertices in $S$ is a 
     satisfying truth assignment for $\Phi$.
\end{proof}

Now, Theorem~\ref{thm:i4l4} follows from Lemma~\ref{lem 4 leaves 4 internal if} and Lemma~\ref{lem 4 leaves 4 internal onlyif}.
A special case of tristar graphs comes as a corollary of Theorem~\ref{thm:i4l4}.

\begin{corollary}
\label{cor:tristar-special}
Let $x,y,z$ be integers such that $1\leq x\leq z$, $y\geq 0$ and either of the following conditions is satisfied.
\begin{enumerate}[label=(\roman*)]
    \item $x=1,y=0,z\geq 3$, or
    \item $x\geq 2$
\end{enumerate}
Then \SCTF{T_{x,y,z}} is \NPC. \TETHS.
\end{corollary}

\subsection{Bistar graphs}
\label{ sec:bistar}
In this section, we prove the hardness for \SCTF{T}, where $T$ is a bistar graph $T_{x,y}$, where $y\geq 5$ and $x\leq y$.
The reduction is from \SCTF{K_{1,y}}.

\begin{theorem}
   \label{thm:bistar}
   Let $x,y$ be two integers such that $1\leq x\leq y$ and $y\geq 5$. Then \SCTF{T_{x,y}}\ is \NPC. \TETHS.
\end{theorem}

\begin{lemma}
\label{lem:star-bistar}
Let $x,y$ be two integers such that $1\leq x \leq y$ and $y\geq 3$. Then
there is a linear reduction from \SCTF{K_{1,y}}\ to \SCTF{T_{x,y}}.
\end{lemma}

Let $T_{x,y}$ be a bistar graph such that $x$ and $y$ satisfy the constraints mentioned in Lemma~\ref{lem:star-bistar}.
Clearly, $T_{x,y}$ has $t = x+y+2$ vertices.
 Construction~\ref{cons star-bistar} is used for the reduction from \SCTF{K_{1,y}}\ to \SCTF{T_{x,y}}.
 
\begin{figure}[ht]
    \centering
    \input{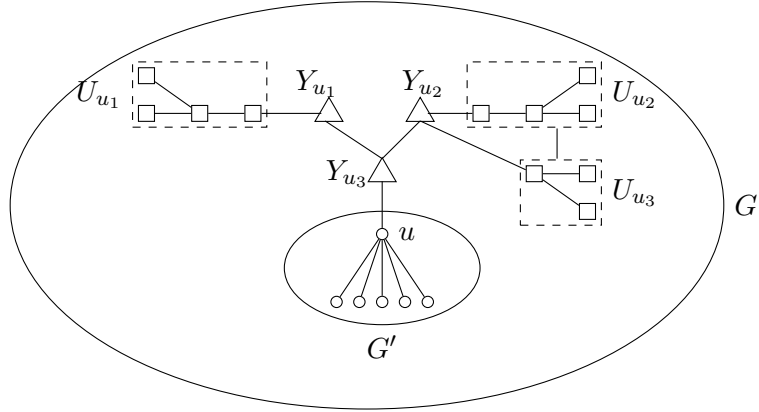}
    \caption{An example of Construction \ref{cons star-bistar} for $x=2$ and $y=5$. 
    Each rectangle (bold) represents a $\overline{T_{x,y}}$ and each triangle represents a $K_5$.  
    The lines connecting two entities (rectangle/triangle/circle) indicate that vertices corresponding to one entity is adjacent to 
    the vertices representing the other entity. 
    } 
    \label{fig: bistar}
\end{figure}

\begin{construction}
\label{cons star-bistar}
    Let $(G',x,y)$ be the input to the construction, where $G'$ is a graph and 
    $x$ and $y$ are integers such that $1\leq x\leq y$ and $y\geq 3$. 
    Let $t= x+y+2$.
    For every vertex $u$ of $G'$,
    introduce $x+1$ sets of $K_y$ denoted by $Y_{u_1},Y_{u_2},\ldots, Y_{u_{x+1}}$, out of which $Y_{u_{x+1}}$ is in the neighbourhood of $u$
    and the sets $Y_{u_1},Y_{u_2},\ldots, Y_{u_x}$ are in the neighbourhood of $Y_{u_{x+1}}$.
    Further, for each set $Y_{u_{i}}$, for $1\leq i\leq x$ introduce a set $U_{u_i}$, which contains 
    $x+2$ sets of $\overline{T_{x,y}}$s denoted by $U_{ij}$ for $1\leq j\leq x+2$.
    The adjacency among these sets $U_{ij}$ and $Y_{u_i}$ is in such a way that taking one vertex from   
    each set $U_{ij}$ along with the complement of $Y_{u_i}$ together induces a $T_{x,y}$.
    Introduce a set of vertices $U_{u_{x+1}}$ which contains $x+1$ copies of $\overline{T_{x,y}}$
    denoted by $U_{(x+1)1}, U_{(x+1)2}, \ldots U_{(x+1)(x+1)}$. The edges from $U_{(x+1)j}$s are in such a way that,
    taking the complement of $Y_{u_{x+1}}$ along with one vertex from $Y_{u_x}$, and one vertex each from $U_{(x+1)j}$s 
    induces a $T_{x,y}$. 
    Further, make $U_{u_x}$ adjacent to $U_{u_{x+1}}$.     
    Let $W_u$ be the set of all new vertices created for a vertex $u\in V(G')$, i.e., 
    $W_u=\bigcup_{i=1}^{i=x+1}(Y_{u_i} \cup U_{u_i})$.
    We note that there are no edges between $W_u$ and $W_{u'}$ for two vertices $u$ and $u'$ in $G'$.
    This completes the construction of the graph $G$ (see Figure~\ref{fig: bistar} for an example).
\end{construction}

The purpose of $U_{u_i}$ is to make sure that not all vertices in $Y_{u_i}$ is in a solution $S$ of $G$, so that if at all 
there is a $K_{1,y}$ induced in $G'\oplus (S\cap V(G'))$, we get a contradiction, as then there will be a $T_{x,y}$ induced in $G\oplus S$
by the vertices in the $K_{1,y}$ and one vertex each, which is not in $S$, from the $Y_{u_i}$s.

\begin{lemma}
\label{lem bistar if}
   If $G' \oplus S' \in \mathcal{F}(K_{1,y})$ for some $S' \subseteq V(G')$, then   $G \oplus S' \in \mathcal{F}(T_{x,y})$.
\end{lemma}

\begin{proof}
   Let a $T_{x,y}$ be induced by a set $A$ in $G\oplus S'$.
   Assume that both the $x$-center (a vertex adjacent to $x$ leaves) $a$ and the $y$-center (a vertex adjacent to $y$ leaves) 
   $b$ of the $T_{x,y}$ are from $G'$.
   Since each vertex in $G'$ is adjacent to only a clique outside $G'$, at most one leaf of $a$  and at most one leaf 
   of $b$   are from outside $G'$. Therefore, $G'\oplus S'$ has an induced $T_{x-1,y-1}$, which contains an induced $K_{1,y}$, a contradiction.
   Let one of the centers, say $u$, be from $V(G')$,  and the other, say $u'$, be from $W_u$. 
   Then $u'\in Y_{u_{x+1}}$.
   Since the size of the maximum independent set in the 
   neighborhood of any vertex in $Y_{u_{x+1}}$ in $W_u$ is $x$, we obtain that there is a $K_{1,y}$ induced in $G'\oplus S'$, which is a contradiction.
   Assume that both $a$ and $b$ are from the new vertices created in $G$. Since $W_u$ and $W_{u'}$ are not adjacent for two vertices $u,u'\in V(G')$,
   we obtain that both $a$ and $b$ are from $W_u$ for some vertex $u\in V(G')$. 
   Let one of the centers, say $v$ of $T_{x,y}$ is from $Y_{u_{x+1}}$. Then the other center, say $v'$ be from any 
   of the sets $Y_{u_j}$ for $1\leq j\leq x$. We observe that
   for every vertex $w\in Y_{u_j}$, the size of the maximum independent set in the neighborhood of 
   $w$ in $W_u\setminus Y_{u_{x+1}}$ is $2 < y$ (recall that each $U_{j\ell}$ induces a $\overline{T}$ which is $3K_1$-free). 
   Therefore, $v'=a$, the $x$-center of $T_{x,y}$, and $v$ is the $y$-center of the $T_{x,y}$.
   Further, $x\leq 2$.
   But, the size of the maximum independent set in the neighborhood of $v$, excluding the clique containing $v'$, is $x$. This implies that $x=y\leq 2$,
   which is a contradiction. 
   Therefore, both $a$ and $b$ are from $W_u\setminus Y_{u_{x+1}}$. 
   It is straight-forward to verify that there are no two adjacent vertices $a,b$ in $W_u\setminus Y_{u_{x+1}}$,
   and an independent set $I$ of size $x+y$ in $W_u\setminus \{a,b\}$ such that $a$ is adjacent to and $b$ is non-adjacent to $x$ vertices in $I$, and
   $b$ is adjacent to and $a$ is non-adjacent to $y$ vertices in $I$. 
\end{proof}

 The converse of the lemma turns out to be true as well. 

\begin{lemma}\label{lem bistar onlyif}
   If $G \oplus S \in \mathcal{F}(T_{x,y})$ for some $S \subseteq V(G)$, then   $G' \oplus S' \in \mathcal{F}(K_{1,y})$, where $S' = S \cap V(G')$. 
\end{lemma}

\begin{proof}
  We observe that for every vertex $u$, set $U_{ij}$ induces a $\overline{T_{x,y}}$. Therefore, $S$ cannot contain all the vertices in $U_{ij}$.
  If $Y_{u_i}$, for $1\leq i\leq x$, is a subset of $S$, then $Y_{u_i}$ and one vertex each from $U_{ij}\setminus S$ (for $1\leq j\leq x+2$) 
  induce a $T_{x,y}$ in $G \oplus S$.
  Therefore, at least one vertex of $Y_{u_i}$ is not in $S$. 
  If $Y_{u_{x+1}}$ is a subset of $S$, then $Y_{u_{x+1}}$ and one vertex from $Y_{u_x}\setminus S$, 
  and one vertex each from $U_{(x+1)j}\setminus S$ (for $1\leq j\leq x+1$) induce
  a $T_{x,y}$. 
  Therefore, at least one vertex of $Y_{u_{x+1}}$ is not in $S$. 
  Assume that there is a $K_{1,y}$ induced by a set $A$ in $G'\oplus S'$. Then, $A$ along with one vertex each 
  from $Y_{u_{j}}\setminus S$, for $1\leq j\leq x+1$, induce a $T_{x,y}$ in $G\oplus S$, which is a contradiction.
\end{proof}

Now, Lemma~\ref{lem:star-bistar} follows from Lemma~\ref{lem bistar if} and Lemma~\ref{lem bistar onlyif}.
Further, Theorem~\ref{thm:bistar} follows from Lemma~\ref{lem:star-bistar} and Proposition~\ref{pro:star}.

\subsection{Tristar graphs}
\label{ sec: tristar graphs}

Recall that, in Section~\ref{sec: 4 leaves 3 internal}, as a corollary (Corollary~\ref{cor:tristar-special}) 
of the main result we have resolved some cases of tristar graphs: 
we proved that \SCTF{T_{x,y,z}} is hard if $z\geq x\geq 2$ or if $x=1,y=0,z\geq 3$.  
In this section, we handle the rest of the cases when $x=1$ and $y\geq 1$, except for a finite number of cases.
First we give a linear reduction from \SCTF{T_{y,z-1}} to \SCTF{T_{x,y,z}}. 
This will take care of the cases when $y\geq 5$ or $z\geq 6$ (recall that the problem for $T_{y,z-1}$ is hard if $y\geq 5$ or $z\geq 6$).
But, for the reduction to work, there is an additional constraint that
$z\geq 3$. So, to handle the case when $z\leq 3$, we introduce another reduction which is from \SCTF{K_{1,y}} 
and does not have any constraint on $z$. Thus, the main result of this section is the following.

\begin{lemma}
   \label{lem:tristar}
   Let $1\leq x\leq z$, and $y\geq 0$ be integers such that $y\geq 5$ or $z\geq 6$.  
   Then \SCTF{T_{x,y,z}} is \NPC. \TETHS.
\end{lemma}

This lemma along with Corollary~\ref{cor:tristar-special} implies the following Theorem.

\begin{theorem}
\label{thm:tristar}
Let $1\leq x\leq z$ and $y\geq 0$ be integers such that at least one of the following conditions is satisfied:
\begin{mylist}
    \item $x\geq 2$, or
    \item $y\geq 5$, or
    \item $z\geq 6$, or
    \item $x=1, y=0, z\geq 3$.
\end{mylist}
Then \SCTF{T_{x,y,z}} is \NPC. \TETHS.
\end{theorem}

First we introduce the reduction from \SCTF{T_{y,z-1}} to \SCTF{T_{x,y,z}}.
\begin{lemma}
   \label{lem:bistar-tristar}
    Let $x,y,z$ be integers such that $x=1, y\geq 1$, and $z\geq 3$.
    Then there is a linear reduction from
    \SCTF{T_{y,z-1}}\ to \SCTF{T_{x,y,z}}.
\end{lemma}
 
 Let $T$ be a tristar graph $T_{x,y,z}$ satisfying the properties stated in Lemma~\ref{lem:bistar-tristar}.
 Construction~\ref{cons bistar-tristar} is used for the reduction from \SCTF{T_{y,z-1}}  to \SCTF{T_{x,y,z}}.
 The reduction is similar to that used for bistars, but simpler.
 
\begin{figure}[ht]
    \centering
    \input{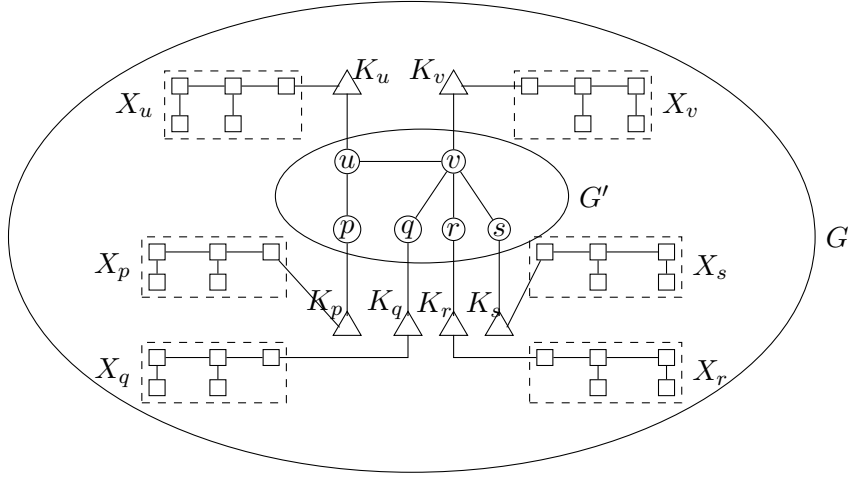}
    \caption{An example of Construction \ref{cons bistar-tristar} with $x=1$ and $y=1$ and $z=4$. 
    Each bold rectangle represents a $\overline{T_{x,y,z}}$ and each triangle represents a $K_4$.  
    The lines connecting two entities (rectangle/triangle/circle) represent all possible edges between the vertices in those entities.}
    \label{fig: tristar-1yz}
\end{figure}

 \begin{construction}
 \label{cons bistar-tristar}
     Let $(G',y,z)$ be the input to the construction, where $G'$ is a graph, $y\geq 1$, and $z\geq 3$ are integers.
     Let $t=y+z+4$ (the number of vertices in $T_{1,y,z}$).
     For every vertex $u$ of $G'$,  introduce a $K_{z}$, denoted by $K_{u}$, in the neighbourhood of $u$.
     Further, introduce $t-z$ copies of  $\overline{T}$ denoted by $X_{u_i}$, for $1\leq i\leq t-z$, in the neighbourhood of $K_{u}$.
     The union of $X_{u_i}$, for $1\leq i\leq t-z$, is denoted by $X_u$.
     The adjacency among these $X_{u_i}$s and $K_u$ is in such a way that taking $z$ vertices from the complement of 
     $K_{u}$ and one vertex each from $X_{u_i}$s induces a $T$.
     Let $W_u$ denote the set of all vertices introduced for a vertex $u$ in $G'$, i.e., $W_u = K_{u}\cup X_{u}$.
     We observe that $W_u$ and $W_v$ are non-adjacent for any two vertices $u,v\in V(G')$.
     This completes the construction of the graph $G$ (see Figure~\ref{fig: tristar-1yz} for an example).
 \end{construction}
 
\begin{lemma}
\label{lem bistar-tristar if}
   If $G' \oplus S' \in \mathcal{F}(T_{y,z-1})$ for some $S' \subseteq V(G')$, then   $G \oplus S' \in \mathcal{F}(T_{x,y,z})$.
\end{lemma}

\begin{proof}
   Let a $T_{x,y,z}$ be induced by a set $A$ in $G\oplus S'$.
   Let $a,b,c$ be the $x$-center, $y$-center, and the $z$-center, respectively of the $T_{x,y,z}$ induced by $A$. 
   Assume that $a,b,c\in V(G')$. 
   Since $a,b$, and $c$ get only at most one leaf from $W_a$, $W_b$, and $W_c$ respectively, there is a $T_{y,z-1}$ in $G'\oplus S'$, which is a 
   contradiction.
   Assume that $a,b\in V(G')$ and $c$ is from $W_b$. 
   Then, $c$ can have at most two leaves from $X_b$, which is a contradiction. 
   The case when $b,c\in V(G')$ and $a\in W_b$ gives a contradiction as there is a $T_{y,z-1}$ in $G'\oplus S'$.
   The case of $a,c\in V(G')$ and $b\notin V(G')$ does not arise as then $a$ and $c$ has a common neighbor outside $V(G')$, which is not true.
   Assume that only $a$ is from $G'$ and $b$, $c$ are from $W_a$. 
   Then $b$ is from $K_u$ and $c$ is from $X_{u_i}$ which is adjacent to $K_u$.
   Then $b$ can have only one leaf adjacent to it which is from $X_{u_i}$. 
   This gives a contradiction as $c$ and a leaf adjacent to $b$ cannot be from the same module.
   The case when $c$ is from $G'$ and $a, b$ are from $W_c$ can be handled in a similar way.
   Note that it is not possible that only $b$, among the centers, is from $G'$, as the neighborhood of $b$
   in $W_b$ is a clique. Assume that $a,b,c\in W_u$ for some vertex $u$ in $G'$. 
   Then $A$ must be subset of $\{u\}\cup K_u\cup X_u$ for some $u\in G'$. 
   It is straight-forward to verify that there is no induced $T_{x,y,z}$ in the graph induced by $\{u\}\cup K_u\cup X_u$
   in $G\oplus S'$.
   This completes the proof.
\end{proof}

The converse of the lemma is also true. 

\begin{lemma}\label{lem bistar-tristar onlyif}
   If $G \oplus S \in \mathcal{F}(T_{x,y,z})$ for some $S \subseteq V(G)$, then   $G' \oplus S' \in \mathcal{F}(T_{y,z-1})$, where $S' = S \cap V(G')$. 
\end{lemma}

\begin{proof}
  Since each $X_{u_i}$ induces a $\overline{T_{x,y,z}}$, at least one of its vertices is not in $S$.
  Therefore, at least one vertex of $K_{u}$ is not in $S$, otherwise, the complement of $K_{u}$
  along with one vertex each from $X_{u_i}\setminus S$ induces a $T_{x,y,z}$.  
  Then, if $G'\oplus S'$ contains a $T_{y,z-1}$ induced by a set $A$ (say), then there is a $T_{x,y,z}$
  in $G\oplus S$ induced by $A$ along with one vertex each from $X_{a}\setminus S, X_b\setminus S$, and $X_c\setminus S$,
  where $b$ and $c$ are the $y$-center and $(z-1)$- center respectively of $T_{y,z-1}$ and $a$ is one of the leaf of $b$ in the $T_{y,z-1}$.
  This completes the proof.
\end{proof}

Lemma~\ref{lem:bistar-tristar} follows from Lemma~\ref{lem bistar-tristar if} and Lemma~\ref{lem bistar-tristar onlyif}.
Now, we introduce the reduction from \SCTF{K_{1,y}} to \SCTF{T_{x,y,z}}.
\begin{lemma}
\label{lem:star-tristar}
    Let $x,y,z$ be integers such that $1\leq x\leq z, y\geq 3$.
    Then there is a linear reduction from
    \SCTF{K_{1,y}}\ to \SCTF{T_{x,y,z}}.
\end{lemma}

Let $T_{x,y,z}$ be a tristar graph satisfying the properties stated in Lemma~\ref{lem:star-tristar}.
 Construction~\ref{cons star-tristar} is used for  the reduction.

\begin{figure}[ht]
    \centering
    \input{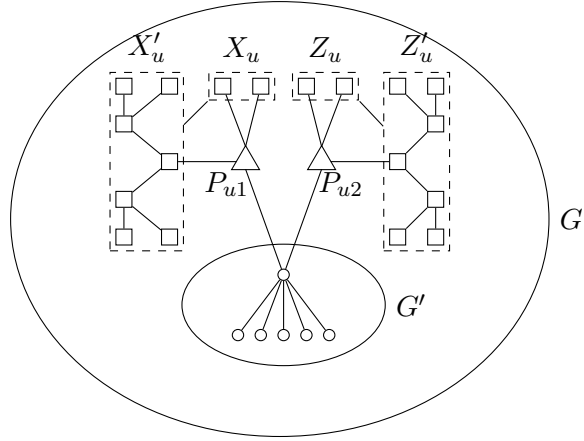}
    \caption{An example of Construction \ref{cons star-tristar} with $x=2$, $y=5$, and $z=2$. 
    Each bold rectangle represents a $\overline{T_{x,y,z}}$ and each triangle represents a $K_5$.  
    The lines connecting two entities (rectangle/triangle/circle) indicate
    the existence of all possible edges between the vertices of the entities.}
    \label{fig: star-tristar}
\end{figure}

\begin{construction}
\label{cons star-tristar}
    Let $(G', x,y,z)$ be the input to the construction, where $G'$ is a graph, and $z\geq x\geq 1$ and $y\geq 3$, are integers.
    Let $t=x+y+z+3$.
    For every vertex $u$ of $G'$,  introduce two $K_{y}$s, denoted by $P_{u1}$ and $P_{u2}$, in the neighbourhood of $u$.
    Further, introduce $x$ copies of  $\overline{T_{x,y,z}}$ denoted by $X_{u_i}$, for $1\leq i\leq x$, in the neighbourhood of $P_{u1}$, 
    and $z$ copies of  $\overline{T_{x,y,z}}$ denoted by $Z_{u_i}$, for $1\leq i\leq z$, in the neighbourhood of $P_{u2}$.
    The union of $X_{u_i}$, for $1\leq i\leq x$, is denoted by $X_u$, and the union of $Z_{u_i}$s, for $1\leq i\leq z$ is denoted by $Z_u$.
    Introduce a set $X_u'$ which contains $t-y$ copies of $\overline{T_{x,y,z}}$s, denoted by $X_{u_i}'$ for $1\leq i\leq t-y$.
    The adjacency among these $X_{u_i}'$s is in such a way that taking $y$ vertices from the complement of $P_{u1}$ and one vertex
    each from $X_{u_i}'$s induces a $T_{x,y,z}$.
    Similarly, introduce a set $Z_{u}'$ which contains $t-y$ copies of $\overline{T_{x,y,z}}$, denoted by $Z_{u_i}'$, for $1\leq i\leq t-y$. 
    The adjacency among these $Z_{u_i}'$s is in such a way that taking $y$ vertices from the complement of $P_{u2}$ and one vertex
    each from $Z_{u_i}'$s induces a $T_{x,y,z}$.
    Further, $X_u$ is adjacent to $X_{u}'$ and $Z_u$ is adjacent to $Z_{u}'$.    
    Let $W_u$ denote the set of all vertices introduced for a vertex $u$ in $G'$, i.e., $W_u = P_{u1}\cup P_{u2}\cup X_u \cup Z_u\cup X_{u}'\cup Z_{u}'$.
    This completes the construction of the graph $G$ (see Figure~\ref{fig: star-tristar} for an example).
\end{construction}

It is straight-forward to verify that the following observation holds true due to the adjacency between $X_u$ and $X_{u}'$, and between $Z_u$ and $Z_u'$.
\begin{observation}
\label{obs:star-tristar}
Let $u$ be any vertex in $G'$. Then there is no $P_4$ induced by vertices in $W_u$ such that at least one of the internal vertex of the $P_4$
is from either $P_{u1}$ or $P_{u2}$.
\end{observation}

\begin{lemma}
\label{lem star-tristar if}
   If $G' \oplus S' \in \mathcal{F}(K_{1,y})$ for some $S' \subseteq V(G')$, then   $G \oplus S' \in \mathcal{F}(T_{x,y,z})$.
\end{lemma}

\begin{proof}
   Let a $T_{x,y,z}$ be induced by a set $A$ in $G\oplus S'$.
   Let $a,b,c$ be the $x$-center, the $y$-center, and the $z$-center, respectively of the $T_{x,y,z}$ induced by $A$. 
   Assume that $a,b,c\in V(G')$. Since $b$ gets only at most two leaves from $W_b$, there is a $K_{1,y}$ in $G'\oplus S'$
   induced by $a,b,c$, and $y-2$ leaves of $b$ from the $T_{x,y,z}$.
   Assume that $a,b\in V(G')$ and $c$ is from $W_b$. Then $b$ can get at most one leaf from $W_b$.
   Therefore, there is a $K_{1,y}$ in $G'\oplus S'$ induced by $a,b$, and $y-1$ leaves of $b$ from the $T_{x,y,z}$.
   The case when $b,c\in V(G')$ and $a\in W_b$ 
   is symmetrical.
   The case of $a$ and $c$ are from $G'$ and $b$ is not from $G'$ does not arise as two vertices in the copy of $G'$ does not have a 
   common neighbor outside $G'$.
   Assume that only $b$ is from $G'$ and $a$ and $c$ are from $W_b$. Then none of the leaves of $b$ are from $W_b$.
   Therefore, there is a $K_{1,y}$ induced by $b$ and its $y$ leaves from the $T_{x,y,z}$.
   Assume that $a\in V(G')$ and $b$ and $c$ are from $W_a$. 
   Then $b$ is from $P_{a1}$ or $P_{a2}$. 
   Then there is a $P_4$ (due to the fact that $y,z\geq 1$) induced by some vertices in $W_a$ such that one of the internal vertex of the $P_4$ is from either 
   $P_{a1}$ or $P_{a2}$. By Observation~\ref{obs:star-tristar}, this is not true.
   Thus we get a contradiction.
   The case when $c\in V(G')$ and $a$ and $b$ are from $W_c$ is symmetrical.
   Therefore, $a,b,c\in W_u$ for some $u\in V(G')$. 
   If $b$ is from $P_{u1}$ or $P_{u2}$, then we get a contradiction using Observation~\ref{obs:star-tristar}. 
   Therefore, $b$ is from $W_u\setminus (P_{u1}\cup P_{u2})$. 
   In this case, it can be verified that, since $y\geq 3$, there is no $T_{x,y,z}$ where at most one leaf is from $G'$.
\end{proof}

The converse of the lemma also is true. 

\begin{lemma}\label{lem star-tristar onlyif}
   If $G \oplus S \in \mathcal{F}(T_{x,y,z})$ for some $S \subseteq V(G)$, then   $G' \oplus S' \in \mathcal{F}(K_{1,y})$, where $S' = S \cap V(G')$. 
\end{lemma}

\begin{proof}
  Since each $X_{u_i}$ induces a $\overline{T_{x,y,z}}$, 
  at least one of its vertices is not in $S$.
  The case is same with $X_{u_i}'$s, $Z_{u_i}$s, and $Z_{u_i}'$s. 
  Therefore, at least one vertex of $P_{u1}$ is not in $S$, otherwise, the complement of $P_{u1}$
  along with one vertex each from $X_{u_i}'\setminus S$ induces a $T_{x,y,z}$. Similarly, $P_{u2}$
  is not a subset of $S$. Then, if $G'\oplus S'$ contains a $K_{1,y}$ induced by a set $A$ (say), then there is a $T_{x,y,z}$
  in $G\oplus S$ induced by $A$ along with one vertex each from $P_{u1}\setminus S, P_{u2}\setminus S$, 
  and one vertex each from $X_{u_i}\setminus S$, for $1\leq i\leq x$,
  and one vertex each from $Z_{u_i}\setminus S$, for $1\leq i\leq z$.
  This completes the proof.
\end{proof}

Lemma~\ref{lem:star-tristar} follows from Lemma~\ref{lem star-tristar if} and Lemma~\ref{lem star-tristar onlyif}.
Now, we are ready to prove Theorem~\ref{thm:tristar}.

\begin{proof}[Proof of Theorem~\ref{thm:tristar}]
Let the integers $x,y,z$ satisfy the constraints given in the theorem, i.e., $1\leq x\leq z$, $y\geq 0$, and either $y\geq 5$ or $z\geq 6$. 
If $x\geq 2$ or if $y=0$, 
then the statements follow from Corollary~\ref{cor:tristar-special}.
Assume that $x=1$ and $y\geq 1$. 
Let $z\geq 6$. Then by Lemma~\ref{lem:bistar-tristar}, there is a linear reduction from \SCTF{T_{y,z-1}} to \SCTF{T_{x,y,z}}.
Then the statements follow from Theorem~\ref{thm:bistar}. 
Let $y\geq 5$. Then by Lemma~\ref{lem:star-tristar}, there is a linear reduction from \SCTF{K_{1,y}} to \SCTF{T_{x,y,z}}.
Then the statement follows from Proposition~\ref{pro:star}.
\end{proof}

\subsection{Paths}
\label{sub:paths}
By Proposition~\ref{pro:path}, \SCTF{P_\ell} is hard for every $\ell\geq 7$.
Here, we extend the result to $P_6$. 

\begin{theorem}
    \label{thm:p6}
    \SCTF{P_6} is \NPC. \TETHS.
\end{theorem}

Proposition~\ref{pro:path} and Theorem~\ref{thm:p6} imply Corollary~\ref{cor:path}.

\begin{corollary}
\label{cor:path}
Let $\ell\geq 6$ be an integer. Then \SCTF{P_\ell} is \NPC. \TETHS.
\end{corollary}

Construction~\ref{cons p6-free} is used for \SCTF{P_6}. The reduction is from \TSAT. The reduction is similar to other reductions that we introduced
from various boolean satisfiability problems. Since a $P_6$ has neither a $4K_1$ nor a $K_4$, the usual technique of keeping an independent set of size 4 
of the literal vertices does not work. To overcome this hurdle, we introduce a vertex  in the clause gadgets.

\begin{figure}[ht]
  \centering
    \input{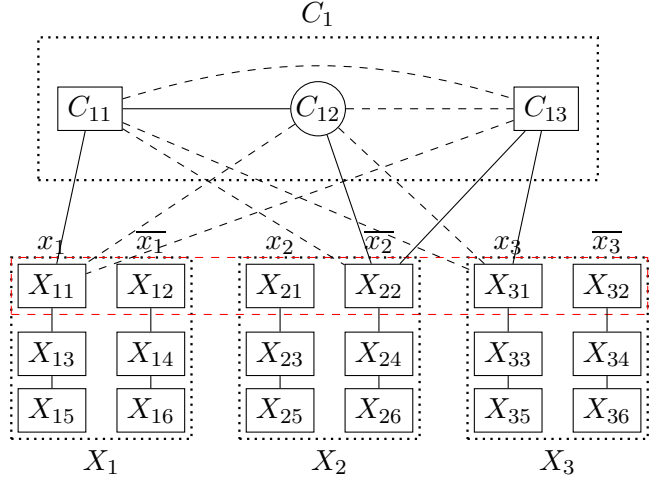}
  \caption{
  An example of Construction \ref{cons p6-free} for the formula $\Phi = (x_1\lor \overline{x_2}\lor x_3)$.
  The bold lines (respectively dashed lines) connecting two entities indicate that each vertex in one entity is adjacent 
  (respectively non-adjacent) to all vertices in the other entity. 
  If there is no line shown between two entities, then the vertices in them are adjacent, with the exceptions --  
  (i) all the vertices in the red rectangle (dashed) together form an independent set; (ii) if there is no line shown 
  between two rectangles in the dotted rectangles, then the rectangles in them are non-adjacent.
  }
    \label{fig:cons P6-free}
  \end{figure}

\begin{construction}
\label{cons p6-free}
    Let $\Phi$ be a \TSAT\ formula with $n$ variables $X_1, X_2, \cdots, X_n$, and $m$ clauses $C_1, C_2, \cdots,$ $C_m$. 
    We construct the graph $G_{\Phi}$ as follows.
    
\begin{itemize}
    \item  For each variable $X_i$ in $\Phi$, the variable gadget, also named  $X_i$, consists of two special sets $X_{i1}=\{x_i\}, X_{i2}=\{\overline{x_i}\}$, and four other sets $X_{i3}, X_{i4},X_{i5}, X_{i6}$, 
    where each of the set in $\{X_{i3}, X_{i4},X_{i5}, X_{i6}\}$ induces a $\overline{P_6}$.
    The set $X_{i1}$ is adjacent to $X_{i3}$ which is adjacent to $X_{i5}$. 
    Similarly, the set $X_{i2}$ is adjacent to $X_{i4}$ which is adjacent to $X_{i6}$. 
    Let $X=\bigcup_{i=1}^{i=n}X_i$. 
    The vertices in $X_{i1} (= \{x_i\})$ and $X_{i2} (=\{\overline{x_i}\})$ (for $1\leq i\leq n-1$)
    are called literal vertices, and let $L$ be the union of all literal vertices.
    The set $L$ induces an independent set of size $2n$.
    
    \item  For each clause $C_i$  of the form  $(\ell_{i1}\lor \ell_{i2}\lor \ell_{i3})$ in $\Phi$, the clause gadget, also named as $C_i$, 
    consists of a set $C_{i2}$ which contains a single vertex $c_{i2}$ and two copies of $\overline{P_6}$ denoted by $C_{i1}$ and $C_{i3}$. 
    The sets $C_{i1}$ and $C_{i2}$ are adjacent.
    Let the three sets introduced (in the previous step) for the literals $\ell_{i1}, \ell_{i2}, \ell_{i3}$ be denoted by  $L_i=\{y_{i1},y_{i2},y_{i3}\}$. 
    Each $C_{ij}$ is adjacent to $y_{ij}$, for $1\leq j\leq 3$, and $C_{i3}$ is adjacent to $y_{i2}$.
    In addition to this, every vertex in $C_i$ is adjacent to all literal vertices corresponding to literals not in 
    $C_i$ with an exception- $C_{i2}$ is adjacent
    to none of the literal vertices corresponding to literals not in $C_i$.
    The union of all clause gadgets $C_i$ is denoted by $C$, and their vertices are called clause vertices.
    
    \item For all $i \neq j$, make the set $C_i$ adjacent to the set $C_j$, and then remove the edge between $C_{i2}$ and $C_{j2}$.
     
    \item For $1\leq i\leq n$, the vertices in $X_{i}\setminus \{x_i,\overline{x_i}\}$ are adjacent to $V(G)\setminus X_i$.  
     
    This completes the construction of the graph $G_{\Phi}$ (see Figure~\ref{fig:cons P6-free} 
    for an example).
   
\end{itemize}
\end{construction}

To prove the forward direction of the correctness of the reduction, we need two lemmas to handle some special cases arising in the forward direction.
\begin{lemma}\label{lem p6-free if - special i}
   Let $\Phi$ be a yes-instance of \TSAT\ and $\psi$ be a truth assignment satisfying $\Phi$.
   Then there exists no set $A$ of vertices such that $A\subseteq (L\cap C)$,
   $(A\cap C)\subseteq (C_i\cup C_j)$, $|A\cap C_i| = 2$, $|A\cap C_j| = 1$, and 
   $A$ induces a $P_6$ in $G_\Phi\oplus S$, where $S$ is the union of the clause vertices $c_{i2}$, 
   for $1\leq i\leq m$, and the set of literal vertices whose corresponding literals were assigned \TRUE\ by $\psi$.
\end{lemma}

\begin{proof}
    Let $A\cap C_i = \{c_{ia}, c_{ib}\}$ and $A\cap C_j = \{c_j\}$.
    Clearly, $c_{ia}c_jc_{ib}$ is a $P_3$.
    Then $A\cap L$ induces either a $P_3$ or a $K_2+K_1$. 
    The former is a contradiction as $L$ induces $K_{n}+nK_1$.
    Assume that $A\cap L$ induces a $K_2+K_1$.
    Let $A\cap L = \{q_1, q_2, q_3\}$ and let $q_2q_3$ be
    the edge in the $K_2+K_1$ induced by $A\cap L$.
    This implies that $q_2,q_3\in S$.
    Note that $c_j$ is adjacent to none of the vertices in $\{q_1,q_2,q_3\}$.
    Since there is no vertex in $C_{\ell1}\cup
    C_{\ell3}$ (for any $1\leq \ell\leq m$) which
    is non-adjacent to three vertices in $L$, we obtain that $c_j\in C_{j2}$.
    Therefore, $c_{j}\in S$.
    Since $c_j$ is in $S$, $c_j$ is non-adjacent to at most one vertex in $S$-- 
    recall that $C_{j2}$ is non-adjacent to all vertices in $L$ except $y_{j2}$, thereby adjacent to all vertices in $S$ except $y_{j2}$ if it is in $S$. 
    Thus, $c_j$ is adjacent to at least one of $q_2,q_3$, which leads to a contradiction.
\end{proof}

\begin{lemma}\label{lem p6-free if - special ii}
   Let $\Phi$ be a yes-instance of \TSAT\ and $\psi$ be a truth assignment satisfying $\Phi$.
   Then there exists no set $A$ of vertices such that $A\subseteq (L\cap C)$,
   $A\cap C\subseteq C_i$ and 
   $A$ induces a $P_6$ in $G_\Phi\oplus S$, where $S$ is the union of the clause vertices $c_{i2}$, 
   for $1\leq i\leq m$, and the set of literal vertices whose corresponding literals were assigned \TRUE\ by $\psi$.
\end{lemma}
\begin{proof}
   If $|A\cap C_i|=1$, then $A\cap L$ induces a graph which has an induced $P_3$, which is a contradiction.
   If $|A\cap C_i|= 2$, then 
   $A\cap C_i$ induces either a $K_2$ or a $2K_1$. 
   Assume that  $A\cap C_i$ induces a $K_2$. 
   Then, $A\cap L$ induces either a $P_4$, or a $P_3+K_1$, or a $2K_2$ - all of them lead to contradictions.
   Assume that  $A\cap C_i$ induces a $2K_1$. 
   Then, $A\cap L$ induces either a $P_4$, or a $P_3+K_1$, or a $2K_2$, or a $K_2+2K_1$ - the first three cases lead to contradictions.
   
   Now assume that $A\cap L$ induces a $K_2+2K_1$.
   Let $c_{ia}$ and $c_{ib}$ be the vertices from $A\cap C$.
   Let $q_1,q_2,q_3,q_4$ be the vertices from $A\cap L$. 
   The vertex $c_{ia}$ is non-adjacent (in $G_\phi\oplus S$) to two vertices in $\{q_1,q_2,q_3,q_4\}$ and the case is same with $c_{ib}$.
   Since $c_{ia}$ and $c_{ib}$ are non-adjacent, one of them must be from $C_{i3}$ (recall that $C_{i1}$ and $C_{i2}$ is adjacent). 
   But vertices in $C_{i3}$ is non-adjacent to exactly one vertex ($y_{i1}$) in $L$.
   This gives a contradiction.

    If $|A\cap C_i|= 3$, then
    $A\cap C_i$ induces a $K_2+K_1$.
    Clearly, $c_{i1}\in C_{i1}$, $c_{i2}\in C_{i2}, c_{i3}\in C_{i3}$
    are the three vertices in $C_i\cap L$,
    where $c_{i1}c_{i2}$ is the edge in the $K_2+K_1$ induced by $A\cap C_i$.
    Then $A\cap L$ induces either a $P_3$ or a $K_2+K_1$ or a $3K_1$. 
    The first case gives a contradiction. 
    Assume that $A\cap L$ induces a $K_2+K_1$. 
    Let $q_1,q_2,q_3$ be the vertices from $A\cap L$ in which $q_1q_2$ is the edge.
    Let $p_1 p_2 p_3 p_4 p_5 p_6$ be the $P_6$ induced by $A$.
    Now there can be four cases- (i)  $p_1, p_2,  p_4 $  belong to $A\cap C_i$ and  $ p_3, p_5, p_6$ belong to $A\cap L$;
    (ii) $p_2, p_3,  p_6 $  belong to $A\cap C_i$ and  $ p_1, p_4, p_5$ belong to $A\cap L$;
    (iii) $p_3,p_4, p_6$ belong to $A\cap C_i$, $p_1,p_2,p_5$ belong to $A\cap L$;
    (iv) $p_1,p_2, p_5$ belong to $A\cap C_i$, $p_3,p_4,p_6$ belong to $A\cap L$.

    Now consider the case (i), i.e.,  $p_1, p_2,  p_4 $  belong to $A\cap C_i$ and  $ p_3, p_5, p_6$ belong to $A\cap L$.
    Clearly, $p_4$ is $c_{i3}$ and $q_2$ and $q_3$ (or equivalently $q_1$ and $q_3$) are adjacent to $c_{i3}$.
    Further, either $c_{i1}$ or $c_{i2}$ is adjacent to $q_3$,
    and both $c_{i1}$ and $c_{i2}$ are non-adjacent to both $q_1$ and $q_2$. But $\{q_1,q_2,c_{i2}\}\in S$.
    This is a contradiction as then $c_{i2}$ must be adjacent to either $q_1$ or $q_2$.
    
    Now consider the case (ii) and (iii): 
    In both cases, $p_6$ must be $c_{i3}$ and hence $c_{i3}$ is non-adjacent to at least two vertices among $\{q_1,q_2,q_3\}$.
    This is a contradiction, as $c_{i3}$ is non-adjacent to only one vertex ($y_{i1}$) in $G_\Phi\oplus S$.

    Now consider case (iv), i.e., $p_1,p_2,p_5$ are in $A\cap C_i$. Then $c_{i3} = p_5$ and either $c_{i1}=p_1$ or $c_{i2}=p_1$.
    If $c_{i2}=p_1$, then $c_{i2}$ is not adjacent to $\{q_1,q_2\}\in S$, which is a contradiction as $c_{i2}$ is non-adjacent to only 
    at most one vertex ($y_{i2}$)
    in $S\cap L$. Therefore, $c_{i1}=p_1$. 
    Then $c_{i1}$ is non-adjacent to all the three vertices $q_1,q_2$, and $q_3$. This is a contradiction as $c_{i1}$ 
    can be non-adjacent to only two vertices ($y_{i2}$ and $y_{i3}$) in $L$. 

    Assume that $A\cap L$ induces a $3K_1$.
    Let $q_1,q_2,q_3$ be the vertices from $A\cap L$.
    Let $p_1 p_2 p_3 p_4 p_5 p_6$ be the $P_6$ induced by $A$. This means that $\{p_2,p_3,p_5\}\in A\cap C_i$, where $p_2p_3$ is an edge.
    Clearly, $c_{i3}$ is $p_5$. There are two cases - either $c_{i1}$ is $p_2$ or $c_{i2}$ is $p_2$. 
    Assume that $c_{i1}$ is $p_2$. Then, without loss of generality, assume that $q_1c_{i1}c_{i2}q_2c_{i3}q_3$ is the $P_6$.
    This means that $p_2$ ($c_{i1}$) is non-adjacent to $q_2,q_3$, which implies that $\{q_2, q_3\}=\{y_{i2}, y_{i3}\}$.
     Similarly, $p_5$ ($c_{i3}$) is non-adjacent to $q_1$, which implies that $q_1=y_{i1}$.
     Thus, $A\cap L\subseteq L_i$. 
     Since $c_{i2}$ and at least one vertex in $L_i$ belong to $S$, $A$ induces a graph which is not isomorphic to $P_6$, which is a contradiction.
     Now assume that $p_3=c_{i1}$, $p_2=c_{i2}$, $p_5=c_{i3}$, $p_1=q_1$, $p_4=q_2$, and $p_6=q_3$. 
     This means that $p_3 = c_{i1}$ is non-adjacent to $q_1,q_3$, which implies that $\{q_1,q_3\}=\{y_{i2},y_{i3}\}$.
     Similarly, $p_5$ ($c_{i3}$) is non-adjacent to $q_1$, which implies that $q_1=y_{i1}$.
     This gives a contradiction as $q_1$ cannot be corresponding two different vertices in $\{y_{i1}, y_{i2}, y_{i3}\}$.
\end{proof}
Now, we prove the forward direction with the help of Lemma~\ref{lem p6-free if - special i} and Lemma~\ref{lem p6-free if - special ii}.
\begin{lemma}\label{lem p6-free if}
   Let $\Phi$ be a yes-instance of \TSAT\ and $\psi$ be a truth assignment satisfying $\Phi$.
   Then $G_{\Phi} \oplus S\in \mathcal{F}(P_6)$ where
   $S$ is the union of the clause vertices $c_{i2}$,
   for $1\leq i\leq m$ and the set 
of literal vertices whose corresponding literals were assigned \TRUE\ by $\psi$.
\end{lemma}

\begin{proof}
    Let $G_{\Phi} \oplus S$ contain a $P_6$ induced by $A$ (say).
    We prove the lemma with the help of a set of claims.
    
    Claim 1: $A$ is not a subset of $X_i$, for $1\leq i\leq n$.

    Assume that $A$ is a subset of $X_i$, for some $1\leq i\leq n$.
    Since $P_6$ is a prime graph and $\overline{P_6}$ is not isomorphic to $P_6$, 
    $A\cap Y\leq 1$, where $Y$ is a module isomorphic to $\overline{P_6}$.
    Thus, $|A\cap X_{ij}| $ is at most one.
    Since $X_i$ has six sets and each of them contains at most one vertex of $A$,  
    each $X_{ij}$ (for $1\leq j\leq 6$) has exactly one vertex of $A$.
    Recall that $\{x_i, \overline{x_i}\}$ is not a subset of $S$.
    Hence, we obtain that the graph induced by $A$ is
    a $2P_3$ which is not isomorphic to $P_6$, which is a contradiction.
    
    Claim 2: For $1\leq i\leq n$, let $X_i'= X_i\setminus \{x_i, \overline{x_i}\}$ and $\overline{X_i} = V(G_\Phi)\setminus X_i$.
    If $|A\cap X_i'|\geq 1$, then $A\cap \overline{X_i}=\emptyset$. Similarly, if $|A\cap \overline{X_i}|\geq 1$,
    then $A\cap X_i'=\emptyset$.
    
    For a contradiction, assume that $A$ contains at least one vertex from $X_i'$ and at least one vertex from $\overline{X_i}$.
    Since $X_i'$ and $\overline{X_i}$ are adjacent, either $|A\cap X_i'|=1$ or $|A\cap \overline{X_i}|=1$.
    
    Assume that $A\cap X_i' = \{u\}$. 
    We note that $V(G_\Phi)\setminus (X_i'\cup \overline{X_i}) = \{x_i, \overline{x_i}\}$.
    Therefore, $|A\cap \overline{X_i}|\geq 3$. This implies that there is a claw ($K_{1,3}$)
    formed by $u$ and three vertices in $\overline{X_i}$, which is a contradiction as there is no claw in a $P_6$.
    Assume that $A\cap \overline{X_i} = \{u\}$. Then with the same argument as given above, we obtain that the 
    graph induced by $A$ contains a claw as a subgraph, which is a contradiction.
    
    Claim 3: $A$ is not a subset of $L$, the set of all literal vertices.
    
    This follows from the fact that $L$ induces a $K_{n}+nK_1$ in $G_\Phi\oplus S$.
    
    Claim 4: $A$ cannot have nonempty intersections with three distinct clause gadgets $C_i$, $C_j$, and $C_\ell$.

    Claim 5: There exists no $C_i$ and $C_j$ ($i\neq j$) such that $|A\cap C_i|\geq 2$ and $|A\cap C_j|\geq 2$.
    
    Claim 4 and 5 follow from the fact that $C_i$ is adjacent to $C_j$ (for $i\neq j$), in $G_\Phi\oplus S$ and the fact that
    there is neither a triangle nor a $C_4$ in a $P_6$.
    
    Claim 6: $A$ is not a subset of $C$.
    
    Since $P_6$ is a prime graph and $\overline{P_6}$ is not isomorphic to $P_6$, $A\cap C_{ij}$ has at most one vertex.
    Therefore, $A\cap C_i$ has at most three vertices.
    By Claim 4, $A$ cannot have nonempty intersections with three clause gadgets $C_i, C_j$ and $C_\ell$.
    Therefore, $A$ has nonempty intersection with exactly two sets $C_i$ and $C_j$ and 
    $|A\cap C_i| = |A\cap C_j| = 3$, which is a contradiction by Claim 5.
    
    Claim 7: If $|A\cap C_i| > 2$, then 
    $A\cap C_j= \emptyset$ ($i\neq j$).
    
    If $A$ contains three vertices from $C_i$ and at least one vertex from $C_j$, then there is a claw in the graph 
    induced by $A$ as $C_i$ and $C_j$ are adjacent in $G_\Phi\oplus S$.

    Now, we are ready to prove the lemma. 
    By Claim 1, $A$ is not a subset of $X_i$ (for $1\leq i\leq n$). 
    By Claim 2, $A$ cannot have vertices from both $X_i\setminus \{x_i, \overline{x_i}\}$ and $\overline{X_i}$ (for $1\leq i\leq n-1$).
    This implies that $A\subseteq L\cup C$. 
    By Claim 3, $A$ cannot be a subset of $L$ and by Claim 6, $A$ cannot be a subset of $C$.
    Therefore, $A$ contains vertices from both $L$ and $C$.
    By Claim 4, $A$ cannot have nonempty intersections with three distinct sets $C_i$, $C_j$ and $C_\ell$.
    Therefore, $A\cap C\subseteq (C_i\cup C_j)$.
    Assume that $A$ has nonempty intersection with both $C_i$ and $C_j$.
    By Claim 5 and Claim 7, we can assume that $|A\cap C_j| = 1$ and $|A\cap C_i|\leq 2$.
    Assume that $|A\cap C_i| = 2$. Then the statement follows from Lemma~\ref{lem p6-free if - special i}.
    Now, assume that $A\cap C_i = \{c_i\}$ and $A\cap C_j = \{c_j\}$. 
    This means that $A\cap L$ induces either a $P_4$, or a $P_3+K_1$, or a $2K_2$ - each of them leads to contradiction. 
    Thus, we can conclude that $A\cap C$ contains vertices from $C_i$ only. 
    Then the statement follows from Lemma~\ref{lem p6-free if - special ii}.
    This completes the proof.
\end{proof}

As usual, the converse is simpler to prove.
\begin{lemma}\label{lem p6-free onlyif}
    Let $\Phi$ be an instance of \TSAT.
    If $G_{\Phi} \oplus S\in \mathcal{F}(P_6)$ for some $S\subseteq V(G_{\Phi})$ then there exists a truth assignment satisfying $\Phi$, i.e.,
    which assigns
    \TRUE\ to at least one literal per clause.
\end{lemma}

\begin{proof}
     Let  $G_{\Phi} \oplus S\in \mathcal{F}(P_6)$ for some $S \subseteq V(G_{\Phi})$. 
     We want to find a satisfying truth assignment of $\Phi$. 
     We know that each of the sets $C_{i1}$ and $C_{i3}$, for $1\leq i\leq m$, induces a $\overline{P_6}$. 
     Therefore, each such set has at least one vertex not in $S$. Hence at least one vertex in $L_i = \{y_{i1}, y_{i2}, y_{i3}\}$ must belong to $S$,
     otherwise there is an induced $P_6$ in $G_\Phi\oplus S$ by vertices in $L_i$ and one vertex each from $C_{i2}$, $C_{i1}\setminus S$, and $C_{i3}\setminus S$.
     
     Similarly, each set $X_{ij}$, for $1\leq i\leq n$ and $3\leq j\leq 6$, induces a $\overline{P_6}$. 
     Therefore, $X_{ij}$ has at least one vertex untouched by $S$. 
     Hence, if both $x_i$ and $\overline{x_i}$
     are in $S$, then there is an induced copy of $P_6$ in $G_\Phi\oplus S$, which is a contradiction.
     Therefore, $\{x_i, \overline{x_i}\}$ is not a subset of $S$. 
     Now, it is straight-forward to verify that assigning \TRUE\ to each literal corresponding to the literal vertices in $S$ is a 
     satisfying truth assignment for $\Phi$.
\end{proof}

Now, Theorem~\ref{thm:p6} follows from Lemma~\ref{lem p6-free if} and Lemma~\ref{lem p6-free onlyif}.

\subsection{Subdivisions of claw}
\label{sub:claw}

A subdivision of a claw has exactly three leaves. Due to this, we cannot handle them using the reduction used to handle trees with 4 leaves
(Theorem~\ref{thm:i4l4}). Let $T = C_{x,y,z}$ be a subdivision of claw, where $x\leq y\leq z$. If $x=y=1$, then $T$ is obtained from 
$P_{z+2}$ by duplicating a leaf. Therefore, we can use Lemma~\ref{lem:duplication} and Corollary~\ref{cor:path} to 
prove the hardness, when $z\geq 4$. If $y>1$, then $T$ is prime and if $T$ has 
at least 9 vertices, then $\overline{T}$ is 5-connected and has an independent set of size 4 (we will prove this in this section). 
Then the hardness results for 5-connected prime graphs (Theorem~\ref{thm:5-conn}) can be used to prove the hardness for $\overline{T}$ and hence for $T$ (Proposition~\ref{pro:complement}).
But, there is a particular subdivision of claw, $C_{1,2,4}$, which is not handled by any of these reductions. Further, there is an infinite family of 
trees, which is obtained by duplicating the leaf adjacent to the center of the claw in $C_{1,2,4}$, not handled by Theorem~\ref{thm:i4l4}, as each tree in the family
violates condition (ii) of Theorem~\ref{thm:i4l4}. 
This requires us to handle $C_{1,2,4}$ separately. We will start this section with a reduction for $C_{1,2,4}$ and end by proving the hardness of \SCTF{T}
if $T$ is not among 7 specific subdivisions of claw.

\begin{figure}[!htbp]
    \centering
    
          \resizebox{0.25\textwidth}{!}{\input{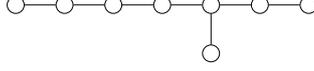} }
     \caption{The subdivided claw $C_{1,2,4}$} 
    \label{fig:eg c124}
\end{figure}

Let $T$ be the subdivided claw $C_{1,2,4}$ shown in Figure \ref{fig:eg c124}. 

\begin{theorem}
    \label{thm:c124}
    \SCTF{C_{1,2,4}} is \NPC. \TETHS.
\end{theorem}

Construction~\ref{cons c124-free} is used for a reduction from \FSAT\ to \SCTF{C_{1,2,4}}. The reduction is similar to the one (Theorem~\ref{thm:i4l4}) 
used to prove hardness 
for trees having 3 internal vertices and 4 leaves.
    
\begin{figure}[ht]
  \centering
    \input{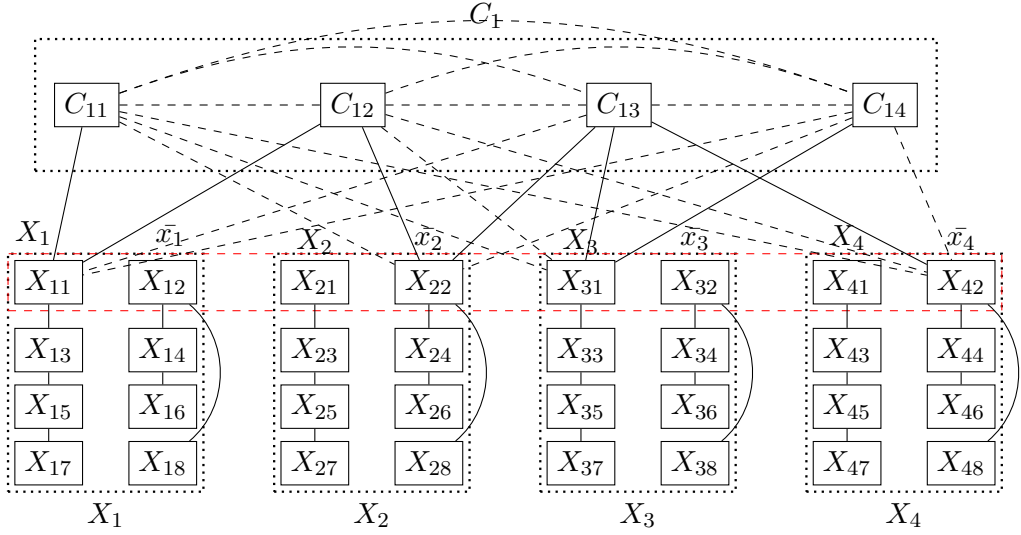}
    \caption{An example of Construction \ref{cons c124-free} for the formula $\Phi = C_1$,  where $C_1=x_1\lor \overline{x}_2\lor x_3\lor \overline{x}_4$. 
    The bold (respectively dashed) lines connecting two rectangles indicate that each vertex in one rectangle is adjacent (respectively non-adjacent) to all vertices in the other rectangle. 
    If there is no line shown between two rectangles, then the vertices in them are adjacent, with an exception --  all the vertices in the red rectangle (dashed) together form an independent set. Similarly,  if there is no line shown between two rectangles in the dotted rectangles, then the rectangles in them are non-adjacent.
   }
  \label{fig:cons c124-free}
\end{figure}

\begin{construction}
\label{cons c124-free}
    Let $\Phi$ be a \FSAT\ formula with $n$ variables $X_1, X_2, \cdots, X_n$, and $m$ clauses $C_1, C_2, \cdots,$ $C_m$. 
    We construct the graph $G_{\Phi}$ as follows.

\begin{itemize}
    \item  For each variable $X_i$ in $\Phi$, the variable gadget, 
    also named  $X_i$, consists of two special sets $X_{i1}=\{x_i\}, X_{i2}=\{\overline{x_i}\}$, and six other sets $X_{i3}, X_{i4},X_{i5}, X_{i6}, X_{i7}, X_{i8}$, 
    where each of the set in $\{X_{i3},X_{i4},X_{i5},  X_{i6},X_{i7},X_{i8}\}$ induces a $\overline{C_{1,2,4}}$.
    The set $X_{ij}$ is adjacent to $X_{i(j+2)}$, for $1\leq j\leq 5$. Further, the sets $X_{i2}$  and $X_{i8}$ are adjacent. 
    Let $X=\bigcup_{i=1}^{i=n}X_i$. 
    The vertices $x_i$ and $\overline{x_i}$ are called literal vertices, and $L$ is the set of all literal vertices.  
    The set $L$ forms an independent set of size $2n$.
    
    \item  For each clause $C_i$  of the form  $(\ell_{i1}\lor \ell_{i2}\lor \ell_{i3}\lor \ell_{i4})$ in $\Phi$, 
    the clause gadget also named as $C_i$ consists of four copies of $\overline{C_{1,2,4}}$s denoted 
    by $C_{i1}$,  $C_{i12}$, $C_{i3}$, and $C_{i4}$. 
    Let the four vertices introduced (in the previous step) for the literals $\ell_{i1}, \ell_{i2}, \ell_{i3}, \ell_{i4}$ be denoted by  $L_i=\{y_{i1},y_{i2},y_{i3}, y_{i4}\}$. 
    The sets $C_{i1}$ and $C_{i2}$ are adjacent to $y_{i1}$. The sets $C_{i2}$ and $C_{i3}$ are adjacent to $y_{i2}$.
    The sets $C_{i3}$ and $C_{i4}$ are adjacent to $y_{i3}$. Additionally, $C_{i3}$ is adjacent to $y_{i4}$.
    Further, every vertex in $C_i$ is adjacent to all literal vertices corresponding to literals not in $C_i$.
    The union of all clause gadgets $C_i$ is denoted by $C$ and their vertices are called clause vertices.
    
    \item For all $i \neq j$, the set $C_i$ is adjacent to the set $C_j$. 
    
    \item  For $1\leq i\leq n$, the vertices in $X_{i}\setminus \{x_i,\overline{x_i}\}$ are adjacent to $V(G)\setminus X_i$.  
     
    This completes the construction of the graph $G_{\Phi}$ (see Figure ~\ref{fig:cons c124-free} 
    for an example)
\end{itemize}
\end{construction}

\begin{lemma}
\label{lem c124-free if}
   Let $\Phi$ be a yes-instance of \FSAT\  and $\psi$ be a truth assignment satisfying $\Phi$. 
   Then $G_{\Phi} \oplus S\in \mathcal{F}(C_{1,2,4})$, where
    $S$ is the set of literal vertices whose corresponding literals were assigned \TRUE\ by $\psi$.
\end{lemma}

\begin{proof}
    Let $G_{\Phi} \oplus S$ contain a $C_{1,2,4}$ 
    induced by $A$ (say).
    We prove the lemma with the help of a set of claims.
    
    Claim 1: $A$ is not a subset of $X_i$, for $1\leq i\leq n$.
    
    Assume that $A$ is a subset of $X_i$.
    Since $C_{1,2,4}$ is a prime graph and $\overline{C_{1,2,4}}$ is not isomorphic to $C_{1,2,4}$, 
    $|A\cap Y|\leq 1$, where $Y$ is a module isomorphic to $\overline{C_{1,2,4}}$.
    Thus, $|A\cap X_{ij}|$ is at most one.
    Therefore, $A$ has nonempty intersection with at least two sets $X_{ij}$ and $X_{i\ell}$.
    Since $C_{1,2,4}$ has 8 vertices, $A$ has nonempty intersection with each set $X_{ij}$ (for $1\leq j\leq 8$).
    Recall that $\{x_i, \overline{x_i}\}$ is not a subset of $S$.
    Hence, we obtain that the graph induced by $A$ is $2P_4$, which is a contradiction.
    
    Claim 2: Let $X_i'= X_i\setminus \{x_i, \overline{x_i}\}$ and $\overline{X_i} = V(G_\Phi)\setminus X_i$.
    If $|A\cap X_i'|\geq 1$, then $A\cap \overline{X_i}=\emptyset$. Similarly, if $|A\cap \overline{X_i}|\geq 1$,
    then $A\cap X_i'=\emptyset$. 
    
    For a contradiction, assume that $A$ contains at least one vertex from $X_i'$ and at least one vertex from $\overline{X_i}$.
    Since $X_i'$ and $\overline{X_i}$ are adjacent, either $|A\cap X_i'|=1$ or $|A\cap \overline{X_i}|=1$.
    Assume that $A\cap X_i' = \{u\}$. 
    Note that $V(G_\Phi)\setminus (X_i'\cup \overline{X_i}) = \{x_i, \overline{x_i}\}$. 
    Therefore, $A$ contains at least 5 vertices from $\overline{X_i}$.
    Then the graph induced by $A$ has a $K_{1,5}$, which is a contradiction, as there is no $K_{1,5}$ in $C_{1,2,4}$.
    Assume that $A\cap \overline{X_i} = \{u\}$. Then with the same argument as given above, we obtain that the 
    graph induced by $A$ contains $K_{1,5}$, which is a contradiction.
    
    Claim 3: $A$ is not a subset of $L$, the set of all literal vertices.
    
    This follows from the fact that $L$ induces a $K_n+nK_1$ in $G_\Phi\oplus S$.
    
    Claim 4: $A$ cannot have nonempty intersections with three distinct clause gadgets $C_i$, $C_j$, and $C_\ell$.

    Claim 5: There exists no $C_i$ and $C_j$ ($i\neq j$) such that $|A\cap C_i|\geq 2$ and $|A\cap C_j|\geq 2$.
    
    Claim 4 and 5 follow from the fact that $C_i$ and $C_j$ are adjacent for $i\neq j$ and $C_{1,2,4}$ does have neither a triangle nor a $C_4$.
    
    Claim 6: $A$ is not a subset of $C$.
    
    For a contradiction, assume that $A\subseteq C$.
    By Claim 4, $A$ cannot have nonempty intersections with three distinct clause gadgets $C_i$, $C_j$, and $C_\ell$.
    Since $C_{1,2,4}$ is a prime graph and $\overline{C_{1,2,4}}$ is not isomorphic to $C_{1,2,4}$, we obtain that $|A\cap C_{ij}|$ is at most one.
    Thus, $A\cap C_i$ induces an independent set of size at most four, which implies that $A$ cannot be a subset of $C_i$.
    Now assume that $A$ has vertices from exactly two sets $C_i$ and $C_j$. 
    Since $C_i$ and $C_j$ are adjacent,  $A$ induces a 
    star graph  which is a contradiction.
    
    Claim 7: If $|A\cap C_i| \geq 2$, then $|A\cap (L\setminus L_i)|$ is at most one.
    
    It follows from the fact that $C_i$  is adjacent to all vertices in $A\cap (L\setminus L_i)$ and $C_{1,2,4}$ does not have  a $C_4$.
    
    We are ready to prove the lemma. 
    By Claim 1, $A$ is not a subset of $X_i$. 
    By Claim 2, $A$ cannot have vertices from both $X_i\setminus \{x_i, \overline{x_i}\}$ and $\overline{X_i}$.
    This implies that $A\subseteq L\cup C$. 
    By Claim 3, $A$ cannot be a subset of $L$ and by Claim 6, $A$ cannot be a subset of $C$.
    Therefore, $A$ contains vertices from both $L$ and $C$.
    By Claim 4, $A$ cannot have nonempty intersections with three distinct clause gadgets $C_i$, $C_j$ and $C_\ell$.
    Therefore, $A\cap C\subseteq (C_i\cup C_j)$.
    Assume that $A$ has nonempty intersection with both $C_i$ and $C_j$.
    By Claim 5, we can assume that $|A\cap C_j| = 1$ and $|A\cap C_i|\geq 1$.
    
    Let $A\cap C_j = \{c_j\}$. 
    Assume that $|A\cap C_i|\geq 4$.
    Then, $A$ induces a graph containing a $K_{1,4}$, which is a contradiction.
    If $|A\cap C_i|=3$, then $A\cap C$ induces a $K_{1,3}$, which implies that $A\cap L$ induces a $P_3+K_1$, which is a contradiction.
    Let $|A\cap C_i|=2$. 
    Since $A\cap C$ induces a $P_3$, $A\cap L$ induces either a $T_{1,2}$, or a $P_4+K_1$, or a $2K_2+K_1$, or a $P_3+P_2$, or a $P_3+2K_1$,
    which is a contradiction.
    Assume that $|A\cap C_i|=1$, then $A\cap C$ induces a $K_2$, which implies that $A\cap L$ induces a graph containing 
    $P_a$, for $a\geq 3$ which is a contradiction. 
    Thus, it is clear that the vertices in $A\cap C$ are from at most one clause gadget $C_i$. 
    
    If $|A\cap C_i|= 1$, then $A\cap L$ induces a graph containing $P_3$. 
    Since  $L$ induces a $K_n+nK_1$, it leads to a contradiction.
    If  $|A\cap C_i|=2$, then also $A\cap L$ induces a graph containing a $P_3$. 
    Let, $|A\cap C_i|=3$. 
    Then by claim 7, $A\cap (L\setminus L_i)$ is at most one.
    Let $A\cap (L\setminus L_i)$ be a singleton set, say $\{w\}$.
    Thus, $A$ contains a $K_{1,3}$ with center as $w$.
    Therefore, $A\cap L_i$ induces a $P_3$ which is a contradiction.
    If  $A\cap (L\setminus L_i)=\emptyset$, then $A$ contains only at most seven vertices, which is a contradiction.
    Now assume that $|A\cap C_i|=4$. 
    Then by claim 7, $A\cap (L\setminus L_i)$ is at most one.
    Let $w\in A\cap (L\setminus L_i)$.
    Thus, the graph induced by $A$ contains a $K_{1,4}$ with center as $w$, which is a contradiction.
    Therefore $A\cap L\subseteq L_i$.
    Since at least two vertices in $A\cap L_i$ is in $S$, the graph induced by $A$ contains at least one edge more than that of $C_{1,2,4}$, which gives a contradiction.
\end{proof}

The backward direction is proved in the next lemma. 

\begin{lemma}\label{lem c124 onlyif}
    Let $\Phi$ be an instance of \FSAT. If $G_{\Phi} \oplus S\in \mathcal{F}(C_{1,2,4})$ for some $S\subseteq V(G_{\Phi})$ then there exists a truth assignment satisfying $\Phi$.
\end{lemma}

\begin{proof}
     Let  $G_{\Phi} \oplus S\in \mathcal{F}(C_{1,2,4})$ for some $S \subseteq V(G_{\Phi})$. 
     We want to find a satisfying truth assignment of $\Phi$. 
     We know that each of the sets $C_{ij}$, for $1\leq i\leq m$ and  $1\leq j\leq 4$, induces a $\overline{C_{1,2,4}}$. 
     Therefore, each such set has at least one vertex not in $S$. Hence at least two vertices in $L_i$ must belong to $S$,
     otherwise there is an induced $C_{1,2,4}$ by vertices in $L_i$ and one vertex each from $C_{ij}\setminus S$, for $1\leq j\leq 4$.
     
     Similarly, each set $X_{ij}$, for $1\leq i\leq n$ and $3\leq j\leq 8$, induces a $\overline{C_{1,2,4}}$. 
     Therefore, $X_{ij}$ has at least one vertex untouched by $S$. 
     Hence, if both $x_i$ and $\overline{x_i}$
     are in $S$, then there is an induced copy of $C_{1,2,4}$ in $G_\Phi\oplus S$, which is a contradiction.
     Therefore, both $\{x_i, \overline{x_i}\}$ is not a subset of $S$. 
     Now, it is straight-forward to verify that assigning \TRUE\ to each literal corresponding to the literal vertices in $S$ is a 
     satisfying truth assignment for $\Phi$.
\end{proof}

Now, Theorem~\ref{thm:c124} follows from Lemma~\ref{lem c124-free if} and Lemma~\ref{lem c124 onlyif}.
We observe that, for any integer $t\geq 4$, the subdivision of claw $C_{1,1,t-2}$ is obtained by introducing a false-twin 
for a leaf of a $P_t$.
Then, Observation~\ref{obs:path-claw} follows directly from Lemma~\ref{lem:duplication}.
\begin{observation}
\label{obs:path-claw}
There is a linear reduction from \SCTF{P_{t}} to \SCTF{C_{1,1,t-2}}.
\end{observation}

Next we prove that $\overline{T}$ is 5-connected for all subdivisions of claw $T$ having at least 9 vertices.
\begin{observation}
\label{obs:claw-5conn}
    Let $T$ be a subdivision of claw. Then $\overline{T}$ is 5-connected if and only if $T$ contains at least 9 vertices.
\end{observation}
\begin{proof}
    It is trivial to observe that the complement of a forest 
    is disconnected if and only if the 
    forest contains a single tree which is a star graph.
    Assume that $T$ has at least 9 vertices. Let $V'$ be a subset of vertices such that $\overline{T}-V'$ is disconnected. 
    Then $T-V'$ is a star graph of at most 4 vertices (there is no star graph of 5 vertices induced in $T$). 
    This implies that $|V'|\geq 5$. Therefore $\overline{T}$ is 5-connected.
    Now, assume that $T$ has only at most 8 vertices. Then, let $V'$ be the set of vertices not in the unique claw in $T$.
    Clearly, $|V'|\leq 4$ and $\overline{T}-V'$ is disconnected. Therefore, $\overline{T}$ is not 5-connected.
\end{proof}

Now, we are ready to prove the main result of this section.

\begin{theorem}
    \label{thm:subclaw}
    Let $x\leq y\leq z$ be integers such that at least one of the following conditions are satisfied.
    \begin{enumerate}[label=(\roman*)]
        \item $x=1,y=2,z=4$, or
        \item $x=y=1$, and $z\geq 4$, or
        \item $x+y+z\geq 8$.
    \end{enumerate}
    Then \SCTF{C_{x,y,z}} is \NPC. \TETHS.
\end{theorem}
\begin{proof}
    Let $T$ be $C_{x,y,z}$.
    If $x=1, y=2,$ and $z=4$, then the statements follow from Theorem~\ref{thm:c124}.
    If $x=y=1$, and $z\geq 4$, then by Observation~\ref{obs:path-claw}, there is a linear reduction from 
    \SCTF{P_{z+2}} to \SCTF{C_{x,y,z}}. Then the statements follow from Corollary~\ref{cor:path}.
    Assume that $y>1$. Then $T$ is a prime graph.
    If $x+y+z\geq 8$,     then $T$ has at least 9 vertices and by Observation~\ref{obs:claw-5conn}, $\overline{T}$ is 5-connected.
    If there is an independent set of size 4 in $T$, then by Theorem~\ref{thm:5-conn}, \SCTF{\overline{T}} is \NPC\ and 
    cannot be solved in subexponential-time, assuming the ETH. Then the statements follow from Proposition~\ref{pro:complement}.
    So, it is sufficient to prove that
    $T$ has an independent set of size 4.
    Let $c$ be the unique vertex with degree 3 in $T$ and let $\{c_1, c_2, c_3\}$ be the leaves in $T$.
    If $\{c,c_1,c_2,c_3\}$ forms an independent set, then we are done. Otherwise, at least one of the leaves, say $c_1$ is adjacent to $c$. 
    Then $T-\{c\}$ contains an isolated vertex $c_1$ and two nontrivial paths such that one of them has length 
    at least two. Then clearly, there is an independent set of size 4 in $T - c$, and hence in $T$.
\end{proof}

Corollary~\ref{cor:subclaw} follows directly from the constraints in Theorem~\ref{thm:subclaw}.
\begin{corollary}
\label{cor:subclaw}
Let $T$ be a subdivision of claw not in $\{$ $C_{1,1,1}, C_{1,1,2}, C_{1,1,3}, C_{1,2,2}, C_{1,2,3}, C_{1,3,3}, C_{2,2,2},$ $ C_{2,2,3}$ $\}$.
Then \SCTF{T} is \NPC. \TETHS.
\end{corollary}
\subsection{Putting them together}
In this section, we prove the main result (Theorem~\ref{thm:tree}) of this paper by using the results proved so far.
We need a few more observations.

\begin{observation}
\label{obs:main-technical}
Let $T$ be a prime tree such that there are two adjacent internal vertices $u,v$ which are not adjacent to any leaf of $T$. 
Then either of the following conditions is satisfied.
\begin{enumerate}[label=(\roman*)]
    \item $T$ has an independent set of size 4 and $\overline{T}$ is 5-connected, or
    \item $T$ is either a $P_6$, or a $P_7$, or the subdivision of claw $C_{1,2,4}$.
\end{enumerate}
\end{observation}
\begin{proof}
Let the neighbor of $u$ other than $v$ be $u'$. Similarly, let the neighbor of $v$ other than $u$ be $v'$.
By the assumption, neither $u'$ nor $v'$ is a leaf. 
Let $T_{u'}$ be the subtree containing $u'$ in $T-u$, and
let $T_{v'}$ be the subtree containing $v'$ in $T-v$.
Let $x$ be the number of vertices in $T_{u'}$ excluding $u'$
and let $y$
be the number of vertices in $T_{v'}$ excluding $v'$, i.e., $x = |T_{u'}|-1$, and $y = |T_{v'}|-1$.
Without loss of generality, assume that $x\leq y$.
Since $u'$ and $v'$ are not leaves, we obtain that $x\geq 1$ and $y\geq 1$.
If $x=1$ and $y=1$, then $T$ is a $P_6$.
If $x=1$ and $y=2$, then $T$ is a $P_7$ (the possibility that $T$ is a $C_{1,1,4}$ does not arise as $T$ is prime).
If $x=1$ and $y=3$, then $T$ is a $C_{1,2,4}$ or a $P_8$ ($T$ is not a $C_{1,1,5}$ as $T$ is prime). But, $P_8$
has an independent set of size 4 and $\overline{P_8}$ is 5-connected. 
If $x=1$ and $y\geq 4$, then there is no subset $V'$ of size at most 4 such that $T-V'$
is a star graph. Therefore, $\overline{T}$ is 5-connected.
Further, there is an independent set of size 4 in $T$.
If $x=2$ and $y=2$, then $T$ is $P_8$ (ignoring the non-prime cases). 
Then $\overline{T}$ is 5-connected and $T$ has an independent set of size 4.
If $x=2$ and $y\geq 3$, then $T$ has an independent set of size 4 and $\overline{T}$ is 5-connected.
The case is same when $x\geq 3$.
\end{proof}

Now, we are ready to prove the main theorem.
\begin{proof}[Proof of Theorem~\ref{thm:tree}]
Let $p$ be the number of internal vertices of $T$.
If $p=1$, then $T$ is a star graph and the statements follow from Proposition~\ref{pro:star}.
If $p=2$, then $T$ is a bistar graph and the statements follow from Theorem~\ref{thm:bistar}.
If $p=3$, then $T$ is a tristar graph and the statements follow from Theorem~\ref{thm:tristar}.
Assume that $p\geq 4$.
If $T$ has only two leaves, then $T$ is isomorphic to $P_\ell$, for $\ell\geq 6$.
Then the statements follow from Corollary~\ref{cor:path}.
If $T$ has exactly three leaves, then $T$ is a subdivision of claw. 
Then the statements follow from Corollary~\ref{cor:subclaw}. 
Assume that $T$ has at least four leaves.

Let $Q_T$ be the quotient tree of $T$.
If $Q_T$ has two adjacent internal vertices which are not adjacent to any leaves, then 
by Observation~\ref{obs:main-technical}, either (i) $Q_T$ has an independent set of size 4 and $\overline{Q_T}$
is 5-connected or (ii) $Q_T$ is either a $P_6$, or a $P_7$, or a $C_{1,2,4}$. 
If (i) is true, then by Theorem \ref{thm:5-conn}, \SCTF{\overline{Q_T}} is \NPC\ and cannot be solved in subexponential-time (assuming the ETH).
Then, so is for \SCTF{Q_T}, by Proposition~\ref{pro:complement}. Then the statements follow from Lemma~\ref{lem:duplication}.
If (ii) is true, then \SCTF{Q_T} is hard by Corollary~\ref{cor:path} and Theorem~\ref{thm:c124}. 
Then the statements follow from Lemma~\ref{lem:duplication}.
Therefore, assume that $Q_T$ has no two adjacent internal vertices not adjacent to any leaves of $Q_T$.
Hence, $T$ has no two adjacent internal vertices not adjacent to any leaves of $T$.
Then, if $T'$, the internal tree of $T$,
is not a star graph, then the statements follow from Theorem~\ref{thm:i4l4}.
Assume that $T'$ is a star graph. 
If the condition (i) of Theorem~\ref{thm:i4l4} is satisfied, then we are done.
Assume that the condition (i) of Theorem~\ref{thm:i4l4} is not satisfied, i.e., the center of $T'$
has at least one leaf of $T$ as a neighbor, one leaf of $T'$ has exactly one leaf of $T$ as a neighbor, and $T$ is neither $C_{1,2,2,2}$ nor $C_{1,2,2,2,2}$.
Assume that $T$ has exactly 4 internal vertices. Then $T'$ is a claw and $Q_T$ is $C_{1,2,2,2}$. Then by Theorem~\ref{thm:i4l4}, 
\SCTF{Q_T} is \NPC\ and cannot be solved in subexponential-time (assuming the ETH). 
Then the statements follow from Corollary~\ref{cor:tree-duplication}. 
Similarly, when $T$ has exactly 5 internal vertices, we obtain that $Q_T$ is $C_{1,2,2,2,2}$ and then the statements follow from 
Theorem~\ref{thm:i4l4} and Corollary~\ref{cor:tree-duplication}. 
Assume that $T$ has at least 6 internal vertices. Then, $T'$ is a $K_{1,a}$, for some $a\geq 5$. Then by Lemma~\ref{lem:leaf-removal}, there is a linear
reduction from \SCTF{K_{1,a}} to \SCTF{Q_T}. By Proposition~\ref{pro:star}, \SCTF{K_{1,a}}
is \NPC\ and cannot be solved in subexponential-time (assuming the ETH). 
Then the statements follow from Corollary~\ref{cor:tree-duplication}. 
\end{proof}

\section{Polynomial-time algorithm}

\label{sec:poly}

The paw is the graph shown in Figure \ref{fig:paw}. 
In this section, we prove that \SCTF{paw} can be solved in polynomial-time. 
We use a result by Olariu~\cite{DBLP:journals/ipl/Olariu88} that
every component of a paw-free graph is either triangle-free or complete mutitpartite.

\begin{figure}[ht]
    \centering
    \input{figs/algo/paw}
    \caption{Paw}
    \label{fig:paw}
\end{figure}

\begin{proposition}[\cite{DBLP:journals/ipl/Olariu88}]
\label{pro:paw-characterization}
A graph $G$ is a paw-free if and only if each component of $G$ is either triangle-free or complete multipartite. 
\end{proposition}

A graph is complete multipartite if and only if it does not contain any $K_2+K_1$ as an induced subgraph. 
It is known that \SCTF{K_3} and \SCTF{K_2+K_1} can be solved in polynomial-time. 
The former is proved in \cite{DBLP:journals/algorithmica/FominGST20} and the latter is implied by another result from \cite{DBLP:journals/algorithmica/FominGST20}
that Subgraph Complementation problems admit polynomial-time algorithms if the target graph class is expressible in MSO\textsubscript{1}
and has bounded clique-width.

\begin{proposition}[\cite{DBLP:journals/algorithmica/FominGST20}]
\label{pro:triangle-cmp}
\SCTF{K_3} and \SCTF{K_2+K_1} are solvable in polynomial-time.
\end{proposition}

Let $G$ be an input graph of \SCTF{paw}. Our algorithm works as follows: First we check whether $G$ can be transformed into a triangle-free graph or a 
complete multipartite graph by Subgraph Complementation, using the algorithms referred in Proposition~\ref{pro:triangle-cmp}.
If yes, then the instance is a yes-instance and we are done. If not, then every solution of $G$ transforms $G$ into a graph having multiple 
components, at least one of it is guaranteed to be a complete multipartite component. 
Then we guess vertices belonging to a 
constant number of those components and then try to obtain $S$ by analysing the neighborhood of the guessed vertices.
We also use the following two observations, which essentially say that it is safe to assume that the input graph does not contain any 
independent module or clique module of size at least 4.

\begin{observation}
\label{obs:paw-is}
Let $G$ be a graph having an independent module 
$I$ of size at least 4. Let $G'$ be the 
graph obtained from $G$ by removing $I$ and introducing an independent module $I'$ of size 3 with the same adjacency as that of $I$. 
Then, $G$ is a yes-instance of \SCTF{paw} if and only if $G'$ is a yes-instance of \SCTF{paw}.
\end{observation}
\begin{proof}
Let $G$ be a yes-instance. Let $S$ be a solution of $G$. Initialize $S' = S\setminus I$. 
Include in $S'$ vertices from $I'$ in such a way that
$|I'\cap S'| = |I\cap S|$, if $0\leq |I\cap S|\leq 1$, and $|I'\cap S'|=2$, if $2\leq |I\cap S|\leq |I|-1$, and
$|I'\cap S'|=3$, if $I\subseteq S$.
We observe that $|I'\cap S'|\leq |I\cap S|$ and $|I'\setminus S'|\leq |I\setminus S|$.
We claim that $S'$ is a solution of $G'$. For a contradiction, assume that there is a set $A'\subseteq V(G')$ which 
induces a paw in $G'\oplus S'$. If $A'$ has no vertices from $I'$, then $A'$ induces a paw in $G\oplus S$, which is a contradiction.
Therefore, $A'\cap I' \neq \emptyset$. 
Initialize $A$ to be $A'\setminus I'$.
Include in $A$, $|I'\cap S'\cap A'|$ vertices from $I\cap S$, and $|(A'\cap I') \setminus S'|$ vertices from $I \setminus S$.
It is straight-forward to verify that this is possible and $A$ induces a paw in $G\oplus S$, which is a contradiction.

For the other direction, assume that $S'$ is a solution of $G'$.
Initialize $S$ to be $S'\setminus I'$. Include vertices from $I$ to $S$ in such a way that $|I\cap S|=|I'\cap S'|$, if $0\leq |I'\cap S'|\leq 2$,
and $|I\cap S|=|I|$ if $I'\subseteq S'$. We claim that $S$ is a solution for $G$.
For a contradiction, assume that $A\subseteq V(G)$ induces a paw in $G\oplus S$.
If $A$ has no vertices from $I$, then $A$ induces a paw in $G'\oplus S'$, which is a contradiction.
Therefore, $A\cap I\neq \emptyset$. 
Initialize $A'$ to be $A\setminus I$.
Since there is no clique module of size three in a paw, we obtain that $|A\cap I\cap S| \leq 2$.
Similarly, since there is no independent module of size two in a paw, we obtain that $|(A\cap I)\setminus S|\leq 1$.
Hence we obtain that $|A\cap I\cap S|\leq |I'\cap S'|$ and $|(A\cap I)\setminus S|\leq |I'\setminus S'|$.
Include in $A'$, $|A\cap I\cap S|$ vertices from $I'\cap S'$, and $|(A\cap I)\setminus S|$ vertices from $I'\setminus S'$.
It is straight-forward to verify that this is possible and $A'$ induces a paw in $G'\oplus S'$, which is a contradiction.
\end{proof}

The proof of Observation~\ref{obs:paw-clique} is similar.
\begin{observation}
\label{obs:paw-clique}
Let $G$ be a graph having a clique module 
$K$ of size at least 4. Let $G'$ be the 
graph obtained from $G$ by removing $K$ and introducing a clique module $K'$ of size 3 with the same adjacency as that of $K$. 
Then, $G$ is a yes-instance of \SCTF{paw} if and only if $G'$ is a yes-instance of \SCTF{paw}.
\end{observation}
\begin{proof}
Let $G$ be a yes-instance. Let $S$ be a solution of $G$. Initialize $S' = S\setminus K$. 
Include in $S'$ vertices from $K'$ in such a way that
$|K'\cap S'| = |K\cap S|$, if $0\leq |K\cap S|\leq 1$, and $|K'\cap S'|=2$, if $2\leq |K\cap S|\leq |K|-1$, and
$|K'\cap S'|=3$, if $K\subseteq S$.
We observe that $|K'\cap S'|\leq |K\cap S|$ and $|K'\setminus S'|\leq |K\setminus S|$.
We claim that $S'$ is a solution of $G'$. For a contradiction, assume that there is a set $A'\subseteq V(G')$ which 
induces a paw in $G'\oplus S'$. If $A'$ has no vertices from $K'$, then $A'$ induces a paw in $G\oplus S$, which is a contradiction.
Therefore, $A'\cap K' \neq \emptyset$. 
Initialize $A$ to be $A'\setminus K'$.
Include in $A$, $|A'\cap K'\cap S'|$ vertices from $K\cap S$, and $|(A'\cap K') \setminus S'|$ vertices from $K \setminus S$.
It is straight-forward to verify that this is possible and $A$ induces a paw in $G\oplus S$, which is a contradiction.

For the other direction, assume that $S'$ is a solution of $G'$. 
Initialize $S$ to be $S'\setminus K'$. Include vertices from $K$ to $S$ in such a way that $|K\cap S|=|K'\cap S'|$, 
if $0\leq |K'\cap S'|\leq 1$, and
$|K\cap S|=|K|-1$ if $|K'\cap S'|=2$, and
$|K\cap S|=|K|$ if $K'\subseteq S'$. We claim that $S$ is a solution for $G$.
For a contradiction, assume that $A\subseteq V(G)$ induces a paw in $G\oplus S$.
If $A$ has no vertices from $K$, then $A$ induces a paw in $G'\oplus S'$, which is a contradiction.
Therefore, $A\cap K\neq \emptyset$. 
Initialize $A'$ to be $A\setminus K$.
Since there is no independent module of size two in a paw, we obtain that $|A\cap K\cap S| \leq 1$.
Similarly, since there is no clique module of size three in a paw, we obtain that $|(A\cap K)\setminus S|\leq 2$.
Hence we obtain that $|A\cap K\cap S|\leq |K'\cap S'|$ and $|(A\cap K)\setminus S| \leq |K'\setminus S'|$.
Include in $A'$, $|A\cap K\cap S|$ vertices from $K'\cap S'$, and $|(A\cap K)\setminus S|$ vertices from $K'\setminus S'$.
It is straight-forward to verify that this is possible and $A'$ induces a paw in $G'\oplus S'$, which is a contradiction.
\end{proof}

It is trivial to note that removing paw-free components from the input graph is safe.
\begin{proposition}
\label{pro:paw:component-removal}
Let $G$ be a graph such that $G'$ is a connected paw-free component of it. Then
$G$ is a yes-instance of \SCTF{paw} if and only if $G - V(G')$ is a yes-instance of \SCTF{paw}.
\end{proposition}

Now onward, we assume that $G$ has no paw-free component and no independent or clique module of size at least 4.
The proposed algorithm for \SCTF{paw} is given below.
Step 1 of our algorithm takes care of the case when $G$ is paw-free. 
If $G$ is not paw-free, then every solution $S$ will have at least two vertices.
Step 2 takes care of the case when there is a solution which transforms the input graph into a single paw-free component.
Step 3 handles the case when there are at least three components in the resultant graph.
Step 4 resolves the case when there are exactly two components.

We define a \textit{component partition} of a graph $G$ as a partition of its vertices into two sets $P,Q$ such that
$P$ induces a single component or an independent set of size at most 3, and $Q$ contains the remaining vertices.
We observe that all component partitions of a graph can be found in polynomial-time.

\vspace{10pt}
\begin{mdframed}
\textbf{Algorithm for \SCTF{paw}}\\
Input: A graph $G$.\\
Output: If $G$ is a yes-instance of \SCTF{paw}, 
then returns \YES;\\
returns \NO~otherwise.
\begin{description}
\item [Step 1]: If $G$ is paw-free, then return \YES.
\item [Step 2]: If $G$ is a yes instance of \SCTF{K_3}, or a yes-instance of \SCTF{K_2+K_1},
then return \YES.
\item [Step 3]: For every triangle $uvw$ in $G$, if $(N(u)\cap N(v)) \cup (N(u)\cap N(w))\cup (N(v)\cap N(w))$ is a solution, then return \YES.
\item [Step 4]: For every ordered pair of adjacent vertices $(u,v)$, do the following:
\begin{enumerate}[label=(\roman*)]
    \item Compute $R_u$ and $R_v$, the lists of component partitions of $N(u)\setminus N[v]$ and $N(v)\setminus N[u]$ respectively.
    \item For every $(X_u, Y_u)$ in $R_u$, and for every $(X_v, Y_v)$ in $R_v$, if $Y_u\cup Y_v\cup (N[u]\cap N[v])$ is a solution, then return \YES.
    \item Let $N_{uv}$ be $N(u)\cap N(v)$.
    \item For every $(X_v,Y_v)$ in $R_v$, and for every subset $S_1'$ of $N_{uv}$ such that $|S_1'|\geq |N_{uv}|-2$, and 
        for every set $V_2'$ of at most three mutually non-adjacent vertices in $G$,
        and for every set $S_2'$ of at most three mutually adjacent vertices in $G$,
        do the following:
        \begin{enumerate}
            \item $S_1'' = Y_v\cup S_1'\cup \{u\}$
            \item $V_2'' = X_v\cup V_2'$
            \item Let $Z_2$ be the set of vertices such that every vertex in $Z_2$ 
            is adjacent to every vertex in $S_1''$ and at least one vertex in $V_2''$.
            \item $S_2'' = S_2'\cup Z_2$
            \item If $S_1''\cup S_2''$ is a solution, then return \YES.
            \item Let $S_2''$ be the set of vertices part of clique components $K$ in the graph $G-S_1''$ such that every vertex in $K$
            is adjacent to every vertex $S_1''$.
            \item If $S_1''\cup S_2''$ is a solution, then return \YES
        \end{enumerate}
\end{enumerate}
\item [Step 5]: Return \NO.
\end{description}
\end{mdframed}
\vspace{10pt}

Next few lemmas state that the algorithm returns \YES\ in various cases.
\begin{lemma}
\label{lem:paw-trivial}
If there exists a set $S\subseteq V(G)$ such that $G\oplus S$ is a connected paw-free graph, then the algorithm returns \YES.
\end{lemma}
\begin{proof}
By Proposition~\ref{pro:paw-characterization}, a connected paw-free graph is either triangle-free or complete multipartite.
Hence $G\oplus S$ is triangle-free or complete multipartite.
Recall that the complete multipartite graphs are exactly the class of $K_2+K_1$-free graphs. 
Then the algorithm returns \YES\ at Step 2.
\end{proof}

Let $S$ be any solution of $G$.
Let $G_1, G_2,\ldots, G_t$, for some integer $t\geq 1$, be the connected components of $G\oplus S$.
Let $S_i$ be the intersection of $S$ with $G_i$ and let  $V_i$ be $V(G_i)\setminus S_i$ (for $1\leq i\leq t$).
For a vertex $u\in S_i$, by $A_u$ we denote the neighbors of $u$ not in $S$. 
Note that $S_i$ and $S_j$ are adjacent in $G$ for $i\neq j$.

\begin{lemma}
\label{lem:paw-3}
Assume that there exists a set $S\subseteq V(G)$ such that $G\oplus S$ is a disjoint union of at least three connected components.
Then the algorithm returns \YES.
\end{lemma}
\begin{proof}
Let $S$ be a set as specified in the lemma. Let $u\in S_1, v\in S_2,$ and $w\in S_3$. 
Note that $N(u)\cap N(v) \subseteq S$ and $S\setminus (S_1\cup S_2)\subseteq N(u)\cap N(v)$. 
Therefore, $(N(u)\cap N(v))\cup (N(u)\cap N(w))\cup (N(v)\cap N(w)) = S$. Then the algorithm returns \YES\ at Step 3. 
\end{proof}

\begin{lemma}
\label{lem:paw-cmp-cmp}
Assume that there exists a set $S\subseteq V(G)$ such that $G\oplus S$ is paw-free and $G\oplus S$ has exactly two components and both of them are 
complete multipartite.
Then
the algorithm returns \YES.
\end{lemma}
\begin{proof}
Let $S$ be a solution as specified in the lemma. 
By Propositon~\ref{pro:paw:component-removal}, we can safely assume $S_1,S_2\neq \emptyset$.
Let $u\in S_1$ and $v\in S_2$. Both $G_1 - S_1$ 
and $G_2 - S_2$ are complete multipartite graphs.
Let $I_1, I_2,\ldots I_{t_1}$ be the independent set partition of $G_1 - S_1$, and let $J_1, J_2,\ldots, J_{t_2}$
be the independent set partition of $G_2 - S_2$. Clearly, each set $I_a$ (for $1\leq a\leq t_1$) 
and $J_a$ (for $1\leq a\leq t_2$) are independent modules in $G$. Therefore, each of them has size at most 3.  
Therefore, if $A_u$ has at least 4 vertices, then $A_u$ induces a connected component in the graph induced by $N(u)\setminus N[v]$.
Further, if $A_u$ induces a disconnected graph, then it forms an independent set of size at most 3, and the vertices of $A_u$ 
are isolated vertices in the graph
induced by $N(u)\setminus N[v]$. 
Similarly, if $A_v$ has at least 4 vertices, then $A_v$ induces a connected component in the graph induced by $N(v)\setminus N[u]$, 
and if $A_v$ induces a disconnected graph, then it forms an independent set of size at most 3, and the vertices of $A_v$ are isolated vertices in the 
graph induced by $N(v)\setminus N[u]$. 
Therefore, $(A_u, (N(u)\setminus N[v])\setminus A_u)$ is a component partition of the graph induced by $N(u)\setminus N[v]$. Similarly,
$(A_v, (N(v)\setminus N[u])\setminus A_v)$ is a component partition of the graph induced by $N(v)\setminus N[v]$.
We note that $(N(u)\setminus N[v])\setminus A_u = S_2\setminus N[v]$ and 
$(N(v)\setminus N[u])\setminus A_v = S_1\setminus N[u]$. 
Therefore, $S = ((N(u)\setminus N[v])\setminus A_u)\cup ((N(v)\setminus N[u])\setminus A_v)\cup (N[u]\cap N[v])$. 
Hence at Step 4(ii), the algorithm returns \YES.
\end{proof}

\begin{lemma}
\label{lem:paw-triangle-cmp}
Assume that there exists a set $S\subseteq V(G)$ such that $G\oplus S$ is paw-free and $G\oplus S$ has exactly two components $G_1$ and $G_2$ such that 
$G_1$ is triangle-free and $G_2$ is complete multipartite.
Then
the algorithm returns \YES.
\end{lemma}
\begin{proof}
Let $u\in S_1$ and $v\in S_2$. Note that $(A_v, (N(v)\setminus N[u])\setminus A_v)$ is a 
component partition of the graph induced by $N(v)\setminus N[u]$ 
($A_v$ induces either a connected graph or has at most three vertices (which forms an independent set)).
Therefore, in some iteration of the loop at Step 4(vi), we get $X_v = A_v$ and $Y_v = S_1\setminus N[u]$.
We recall that there are no clique module of size at least 4. 
Note that $N[v]\cap S_2$ is a clique module. Therefore, only at most two vertices of $N_{uv} (= N(u)\cap N(v))$ are not in $S_1$.
Therefore, in one iteration of the loop at Step 4(vi), we obtain that $S_1'= N(u)\cap S_1$.
Hence in Step 4(vi)(a), we obtain that $S_1'' = S_1$.
Since there are no independent module of size at least 4. 
Therefore, $X_v$ has all the vertices of $G_2-S_2$, except possibly at most three mutually non-adjacent vertices.
We obtain that set $V_2'$ in one iteration of the loop at Step 4(vi). 
Therefore, in Step 4(vi)(b), we obtain $V_2'' = V(G_2)\setminus S_2$. 
Let $p$ be the number of parts in the partition of the complete multipartite graph $G_2-S_2$.

If $p\geq 2$, then every vertex in $S_2$ is adjacent to at least one vertex in $G_2-S_2$. 
Therefore, we obtain $S_2'' = S_2$ in Step 4(vi)(d) - for an iteration of Step 4(vi) in which $S_2'$ is empty.
Therefore, the algorithm return \YES\ in Step 4(vi)(e).

Now, assume that $p=1$. Then every vertex in $S_2$, except possibly at most three mutually adjacent vertices, is adjacent to at least one 
vertex in $G_2-S_2$. Therefore, in Step 4(vi)(d), we obtain that $S_2 = S_2''$ - for an iteration of Step 4(vi) in which $S_2'$ is the 
set of at most three mutually adjacent vertices in $S_2$ non-adjacent to any vertex in $G_2-S_2$.
Hence, the algorithm returns \YES\ in Step 4(vi)(e).

Assume that $p=0$, i.e., $G_2-S_2$ has no vertices.
Then we can construct a solution $W = W_1\cup W_2$ such that $W_1 = S_1$ and $W_2$ is exactly the set of vertices part of the clique components 
$K$ of $G-S_1$ such that every vertex in $K$ is adjacent to 
all vertices in $S_1$. Then the algorithm returns \YES\ in Step 4(vi)(g).
\end{proof}

\begin{lemma}
\label{lem:paw}
The algorithm returns \YES\ if and only if $G$ is a yes-instance.
\end{lemma}
\begin{proof}
Clearly, if the algorithm returns \YES, then $G$ is a yes-instance. For the other direction, assume that $G$
is a yes-instance. 
If there exists a solution $S$ such that $G\oplus S$ is a connected component or a triangle-free graph, 
then the algorithm returns \YES\ by 
Lemma~\ref{lem:paw-trivial}. 
If there exists a solution $S$ such that $G\oplus S$ has at least three connected components, then by Lemma~\ref{lem:paw-3},
the algorithm returns \YES.
Assume that there is no solution that transforms the graph into a single component or at least three components.
Further assume that there is no solution that transforms the graph into a triangle-free graph.
Then there must exists a solution $S$ such that $G\oplus S$ has exactly two connected components $G_1$ and $G_2$, such that one of them, say $G_2$, is a 
complete multipartite graph.
Assume that $G_1$ is also complete multipartite. Then by Lemma~\ref{lem:paw-cmp-cmp}, the algorithm returns \YES.
Now, assume that $G_1$ is triangle-free graph. Then the algorithm returns \YES\ by Lemma~\ref{lem:paw-triangle-cmp}.
This completes the proof.
\end{proof}

Now, Theorem~\ref{thm:paw} follows from Lemma~\ref{lem:paw} and the fact that all split partitions and all 
component partitions of a graph can be obtained in polynomial-time.
\begin{theorem}
\label{thm:paw}
\SCTF{paw} can be solved in polynomial-time. 
\end{theorem}

\section{Concluding remarks}
\label{sec:conclusion}

In this paper, we resolved the computational complexity of \SCTF{T}, for all trees $T$, except for 37 trees listed in Figure~\ref{table:opencases}.
Among these open cases, we would like to highlight the tree $C_{1,2,2}$. If we can prove that \SCTF{C_{1,2,2}} is hard, then the list of open 
cases reduces to 17 trees, i.e., all the trees numbered 18 to 37 in the list vanishes due to Corollary~\ref{cor:tree-duplication}. 
The tree resisted all our attempts to cut it down. Among other open cases, we believe that those with 5 vertices ($P_5, K_{1,4}$, 
and $T_{1,2}$) are the most challenging - we do not have any result so far on non-trivial 5-vertex graphs. 
The case of $P_5$ was stated as an open problem in \cite{DBLP:journals/algorithmica/FominGST20}.
We also believe 
that the claw may admit a polynomial-time algorithm, similar to paw - 
the difficulty in getting such a result seems to reside in the intricacies of the structure theorem for claws. 

To get a complete P versus \NPC\ dichotomy for \SCTF{H}, for general graphs $H$, one major hurdle is to tackle the graphs which are self-complementary. 
Introducing $\overline{H}$ in a reduction for \SCTF{H} helps us to make sure that at least one vertex in a set of vertices is untouched by any solution. 
We find it very difficult to find alternate reductions which do not use $\overline{H}$ -- a reason 
why we do not have hardness results so far for any self-complementary graphs.

\bibliographystyle{unsrt}
\bibliography{main}
\end{document}